%% file: paper.tex
\documentclass[11pt,a4paper]{article}

\pdfoutput=1
\usepackage{amssymb}
\usepackage{amsmath}
\usepackage{amsfonts}
\usepackage{graphicx}
\usepackage{booktabs}
\usepackage{color}
\usepackage{textcomp}
\usepackage{multirow}
\usepackage{bm}
\usepackage{caption}
\usepackage{subcaption}
\usepackage{tikz}
\usepackage[export]{adjustbox}
\usepackage{longtable}
\usepackage{colortbl}
\usepackage{rotating}
\usepackage{jheppub}
\usepackage[utf8]{inputenc}
\usepackage{braket}
\usepackage{hyperref}
\usepackage{comment}
\usepackage{mathtools}
\usepackage{epstopdf}
\usepackage[labelfont=bf]{caption}
\usepackage[section]{placeins}
\usepackage{bbold}
\usepackage[colorinlistoftodos]{todonotes}

\definecolor{notenic}{rgb}{0.61, 0.77, 0.89}
\definecolor{notedave}{rgb}{1.0, 0.75, 0.0}

\definecolor{jlab_red}{RGB}{192,39,45}
\definecolor{jlab_orange}{RGB}{249,102,0}
\definecolor{jlab_blue}{RGB}{47,122,121}
\definecolor{jlab_green}{RGB}{65,125,10}

\definecolor{swave}{HTML}{B23B3B}
\definecolor{pwave}{HTML}{424874}
\definecolor{dstarpi}{HTML}{A15017} 

\definecolor{dstarpi_s}{RGB}{191,39,45}
\definecolor{dstarpi_d}{RGB}{65,125,10}
\definecolor{dstarpi_mix}{RGB}{248,102,0}
\definecolor{dpi_d}{RGB}{51,92,129}
\definecolor{dstareta_s}{HTML}{A647FF}
\definecolor{dsstarkbar_s}{HTML}{258EA6}
\definecolor{dstareta_dstarpi_s}{HTML}{b33796}
\definecolor{dsstarkbar_dstareta_s}{HTML}{666bd3}
\definecolor{dsstarkbar_dstarpi_s}{HTML}{8c4955}

\usepackage{tocloft}
\newcommand{\cm}{\ensuremath{\mathsf{cm}}}

\newcommand{\Ndf}{\ensuremath{N_\mathrm{dof}}}

\newcommand{\thr}{\ensuremath{\text{thr}}}
\renewcommand{\Re}{\ensuremath{\text{Re }}}
\renewcommand{\Im}{\ensuremath{\text{Im }}}

\hypersetup{%
pdftitle = {D1 and D2 resonances in coupled-channel scattering amplitudes from lattice QCD},
pdfkeywords = {lattice, QCD, meson, hadron, scattering, charm},
pdfauthor = {Hadron Spectrum Collaboration},
colorlinks = {true},
filecolor = {black},
linkcolor = {jlab_blue},
menucolor = {black},
citecolor = {jlab_green},
urlcolor = {jlab_green},
}{}

\newcommand{\dpi}{\ensuremath{{D\pi}}}
\newcommand{\deta}{\ensuremath{{D\eta}}}
\newcommand{\dskbar}{\ensuremath{{D_s\bar{K}}}}
\newcommand{\dstarpi}{\ensuremath{{D^\ast \pi}}}
\newcommand{\dstareta}{\ensuremath{{D^\ast \eta}}}
\newcommand{\dsstarkbar}{\ensuremath{{D_s^\ast \bar{K}}}}

\newcommand{\dzerostar}{\ensuremath{{D_0^\ast}}}
\newcommand{\dszerostar}{\ensuremath{{D_{s0}^\ast}}}

\newcommand{\SLJ}[3]{\ensuremath{{\:\!}^{{#1}\!}{#2}_{#3}}}
\newcommand{\SLJc}[3]{\ensuremath{ { \{  {\:\!}^{{#1}\!}{#2}_{#3}  \}  } } }

\newcommand{\osz}{\ensuremath{\SLJ{1}{S}{0}}}
\newcommand{\tso}{\ensuremath{\SLJ{3}{S}{1}}}
\newcommand{\tpz}{\ensuremath{\SLJ{3}{P}{0}}}
\newcommand{\opo}{\ensuremath{\SLJ{1}{P}{1}}}
\newcommand{\tpo}{\ensuremath{\SLJ{3}{P}{1}}}
\newcommand{\tpt}{\ensuremath{\SLJ{3}{P}{2}}}
\newcommand{\tdo}{\ensuremath{\SLJ{3}{D}{1}}}
\newcommand{\odt}{\ensuremath{\SLJ{1}{D}{2}}}
\newcommand{\tdt}{\ensuremath{\SLJ{3}{D}{2}}}
\newcommand{\tdthr}{\ensuremath{\SLJ{3}{D}{3}}}
\newcommand{\tgt}{\ensuremath{\SLJ{3}{G}{3}}}

\begin{document}

\input{title}
\maketitle

\input{intro}

\input{spectra}

\input{scattering}

\input{poles}

\input{interpretation}

\input{summary}

\bigskip

\input{ack}

\newpage
\appendix
\section*{Appendices}

\input{appendix_operators}

\input{appendix_amp_variations}

\bibliography{biblio}
\bibliographystyle{JHEP}

\end{document}

%% file: title.tex

\title{$D_1$ and $D_2$ resonances in coupled-channel scattering amplitudes from lattice QCD}

\author[a]{Nicolas Lang}
\author[b]{and David J. Wilson}
\author{\\for the Hadron Spectrum Collaboration}

\affiliation[a]{Instituto de F\'{i}sica Corpuscular (IFIC), Universitat de Valencia, 46071, Valencia, Spain}
\affiliation[b]{Department of Applied Mathematics and Theoretical Physics, Centre for Mathematical Sciences,\\ University of Cambridge, Wilberforce Road, Cambridge, CB3 0WA, UK}

\emailAdd{nicolas.lang@ific.uv.es}
\emailAdd{d.j.wilson@damtp.cam.ac.uk}

\abstract{
Isospin-1/2 charmed axial-vector $D^*\pi-D^*\eta-D^*_s\bar{K}$ scattering amplitudes are computed, along with interactions in several other $I=1/2$ $J^P$ channels. Using lattice QCD, we work at a light-quark mass corresponding to $m_\pi\approx 391$ MeV, where the lowest three-hadron threshold ($D\pi\pi$) lies high enough to enable a rigorous treatment of this system considering only two-hadron scattering channels. At this light-quark mass, an axial-vector $D_1$ bound state is observed just below $D^\ast\pi$ threshold, that is strongly coupled to $D^\ast\pi$ in a relative $S$-wave and influences a wide energy region up to the $D^\ast\eta$ threshold. An axial-vector $D_1^\prime$ resonance is observed in the elastic $D^\ast\pi$ energy-region, which is coupled more strongly to $D$-wave $D^\ast\pi$. A single narrow tensor state is seen in $J^P=2^+$ coupled to both $D\pi$ and $D^\ast\pi$. In the region where $D^*\eta$ and $D^*_s\bar{K}$ are kinematically open, the available energy levels indicate significant $S$-wave interactions. Upon searching this region for poles, several possibilities exist with large uncertainties. One additional state consistently arises,  predominantly coupled to the $S$-wave $D^*\pi-D^*\eta-D^*_s\bar{K}$ amplitudes around the upper energy limit of this analysis.
}

%% file: intro.tex
\section{Introduction}
\label{sec:intro}

An era of renewed interest in hadron spectroscopy has been ushered in by the experimental discovery of a large number of excited-state hadrons, many of which have valence charm-quark contributions. These excited states are seen as resonant enhancements in the event rates of lighter hadrons and are thus classified as unstable resonances. Some have exotic quantum numbers and entirely evade conventional quark-model descriptions. Others may be compatible with a quark-model construction in terms of their flavour content, but their appearance in the spectrum and couplings to strong-decay channels require more than this simple picture.

One set of states that has garnered significant attention is that of the lightest open-charm mesons. These are states with a minimal $c\bar{q}$ valence-quark content, where $q$ could either be a light or a strange quark. The ground-state pseudoscalar $D$ and vector $D^*$ mesons ($D_s$ and $D_s^*$ in the charm-strange sector) are well described as quark-model $S$ wave systems. In this framework, the $\dzerostar(2300)$, $D_1(2430)$, $D_1(2420)$ and $D_2^*(2460)$ could be collectively identified with the first orbital excitation of the $c\bar{q}$ system. But this interpretation has been increasingly challenged. One reason is that the $\dzerostar(2300)$ mass was measured to be equal or even larger than that of the $\dszerostar (2317)$, contrary to the expectation that the mass gap between the light and the strange quark should be the primary contribution to their mass difference. Furthermore the $\dzerostar(2300)$ and $D_{1}(2430)$ have signatures in their decay amplitudes that are very different from those of their strange siblings. They are broad features above the relevant strong-decay thresholds, $\dpi$ and $\dstarpi$ respectively, whereas the charm-strange states are narrow resonances below the $D^{(*)}K$ thresholds.

These observations call for more elaborate tools than quark-potential models. A rigorous description of unstable particle states is in terms of pole singularities in the scattering amplitude, which, apart from these singularities and branch cuts, is considered an analytic function of energy on the complex plane. Poles near the real axis and distant from any decay thresholds produce a simple signature in the amplitude on the real axis, corresponding to an approximately symmetric peak that resembles the shape of a Breit-Wigner parametrisation. The narrow $D_1(2420)$ and $D_2^*(2460)$ are examples for this type of resonance. Large couplings to strong-decay thresholds in the vicinity of the resonance can substantially alter this picture. In the case of the $\dzerostar(2300)$ it appears likely that the pole singularity and the amplitude peak differ significantly in energy~\cite{Du:2020pui,gayer:2021}. 
This suggests a $\dzerostar(2300)$ located at a much lower mass than what is currently reported by the \emph{Particle Data Group} (PDG)~\cite{ParticleDataGroup:2024}, restoring the natural mass order between the lightest scalar charm-light and charm-strange states. 

A strong coupling to the respective threshold furthermore invalidates a simple interpretation of these as $c\bar{q}$ states. Even if a mainly $c\bar{q}$-like configuration is assumed, the presence of meson-meson decay modes necessitates a meson-meson component. Above threshold, large couplings to these decay modes can push the pole far into the complex plane as the phase space opens. Beyond $c\bar{q}$-like and meson-meson-like configurations, compact tetraquark objects have also been suggested~\cite{Maiani:2004vq}. To further complicate matters amplitudes can have multiple singularities whose effects overlap. They can also interact with multiple scattering thresholds that emerge
whenever the same flavour quantum numbers can be formed from different hadron-hadron combinations.

In QCD, an SU(3) flavour symmetry arises when the up, down and strange quark masses are set equal. This symmetry is broken in nature by the differing quark masses, however it appears relevant for relating many processes. For an anti-triplet $D$-meson interacting with an octet or single light meson, the relevant flavour decomposition is $\bm{\overline{3}}\otimes {{\bm 8}} \to {\overline{\bm 3}}\oplus {{\bm 6}}\oplus \overline{\bm{15}}$. The ${\bm 6}$ and $\overline{\bm{15}}$ can only be formed from ``exotic'' 4-quark combinations while the $\bar{\bm 3}$ mixes with 4-quark and $c\bar{q}$ combinations. In the SU(3) limit, the interactions within each flavour irreducible representation (irrep) are identical. Thus if SU(3) symmetry is not too badly broken in nature, then this should be apparent in the interactions of $D/D_s$-mesons with light and strange mesons.

By considering the three-flavour variant of chiral perturbation theory~\cite{Gasser:1983yg,Gasser:1984gg} applied when one heavy quark is present~\cite{Jenkins:1990jv,Yan:1992gz,Wise:1992hn} and expanding its range of applicability by ``unitarising''~\cite{Truong:1988zp,Oller:1997ti} (SU(3) u$\chi_{\text{PT}}$), it is possible to consider the interactions of charmed mesons ($D^{(*)},D_s^{(*)}$) in a flavour triplet with the octet of light and strange pseudoscalar mesons ($\pi$, $K$, $\eta$)~\cite{Kolomeitsev:2003ac}.  It is observed that the anti-triplet is strongly attractive, containing near-threshold bound states and resonances, the sextet is also attractive, while the $\overline{\bm{15}}$ is weakly repulsive.
Using SU(3) u$\chi_{\text{PT}}$, several studies assert that there is a pole in the $S$-wave sextet amplitude~\cite{Kolomeitsev:2003ac,Hofmann:2003je,Gamermann:2006nm,Guo:2006fu,Guo:2009ct,Doring:2011vk,ALBALADEJO2017465,Du:2017zvv,Meissner:2020khl} which would affect any flavour combinations that contain this irrep. Among these are the non-strange, isospin-1/2 amplitudes\footnote{But also simpler cases such as $I=0$ $D\bar{K}$}, relevant for $D^{(*)}\pi$, $D^{(*)}\eta$ and $D_s^{(*)}\bar{K}$ scattering. As the number of poles tends to be relatively stable as a function of the quark masses, it offers a possible explanation for some of the similarities and differences between the strange and non-strange $D$-meson scattering amplitudes, in particular at higher energies.

Ultimately all of these states and their interactions originate in QCD and are the consequences of the non-perturbative dynamics between quarks and gluons. Unfortunately, low-energy QCD is neither easy to treat mathematically nor easy to interpret phenomenologically, which is why models and effective field-theory approximations have an important role. However, they rely on external inputs that have to come either from experimental data, or from calculations utilising QCD itself.
The only tool known to-date that allows to make non-perturbative QCD predictions with controlled systematic uncertainties is lattice QCD. In conjunction with a formalism that maps finite-volume energy spectra to infinite-volume scattering amplitudes, lattice QCD enables the computation of strong scattering processes.
It has the benefit that all possible channels that contribute in a given system can be considered, thus building a ``complete'' $S$-matrix that satisfies the principles of scattering theory such as unitarity and analyticity.

In this work, we determine lattice QCD finite-volume spectra with charm $C=1$, strangeness $S=0$, and isospin $I=\frac{1}{2}$ quantum numbers from correlation matrices computed using a large basis of operators. These energy levels are used to constrain coupled $J^P=1^+$ $\dstarpi$, $\dstareta$ and $\dsstarkbar$ and coupled $J^P=2^+$ $D\pi$ and $\dstarpi$ scattering amplitudes which we investigate for their singularity content. 
We will find two low-lying axial-vector D-meson states and a tensor resonance, which we relate to the $D_1(2430)$, $D_1(2420)$ and $D_2^*(2460)$. The pole corresponding to the $D_1(2430)$ appears to lie below the PDG-reported value, similar to the scalar sector. We will also find at least one additional axial-vector state above the $\dsstarkbar$ scattering threshold, which may be related to the conjectured sextet-pole.
This work is an extension of our earlier article ref.~\cite{Lang:2022elg}, where we first constrained the elastic $\dstarpi$ amplitude in the axial-vector channel; here we include more energy levels in our fits and attempt to constrain higher energies, including further hadron-hadron channels.
In ref.~\cite{Lang:2022elg}, heavy-quark spin-symmetry appeared to be reasonably well satisfied and we consider this again working at higher energies.
This work also complements ref.~\cite{Moir:2016srx}, which studied coupled $\dpi$, $\deta$ and $\dskbar$ scattering, investigating the scalar $\dzerostar(2300)$ using the same lattice ensembles, as well as ref.~\cite{gayer:2021}, studying $\dpi$ scattering at a lighter pion mass. Furthermore, the results from ref.~\cite{Cheung:2020mql} considering $DK$ scattering with the same lattices, and ref.~\cite{Yeo:2024chk} considering $D/D_s$ mesons with SU(3) flavour symmetry are relevant to this analysis, in particular to the discussion of the sextet pole.
Other lattice QCD studies have considered the interactions of $D$ and $D_s$-mesons~\cite{Liu:2012zya,Mohler:2013rwa,Lang:2014yfa,Bali:2017pdv,Alexandrou:2019tmk}, however the coupled-channel region remains relatively unexplored.

The paper is organised as follows: In section~\ref{sec:lattice} we present the lattice computation and finite-volume spectra. In section~\ref{sec:scattering} we fit scattering amplitudes to the spectra. In section~\ref{sec:poles} we analytically continue the amplitudes to the complex plane and search for pole singularities. In section~\ref{sec:interpretation} we interpret our results. Section~\ref{sec:summary} contains our summary and conclusions.

%% file: spectra.tex

\section{Lattice calculation}
\label{sec:lattice}

In this section we present the finite-volume spectra, which are the input for the scattering analysis, and briefly explain how they are obtained from a lattice calculation. We begin by outlining the important features of the  $I=1/2$, $C=1$ $S=0$ meson-meson system as found in a continuous, infinite spacetime, before moving on to the technicalities of the lattice and the results of the calculation.

\subsection{Charm-light scattering in the continuum}

Our primary aim in this work is to study unnatural parity $J^P=1^+$, which is found in $S$-wave vector-pseudoscalar scattering. We use unphysically massive light quarks that render the $D^\ast$ absolutely stable and raise three-hadron thresholds well-above the energy region we intend to study. The relevant two-meson combinations are $D^\ast\pi$, $D^\ast\eta$, $D_s^*\bar{K}$ and $D^\ast\eta^\prime$. The lowest three-hadron channels are $\{D,D^\ast\}-\pi\pi$ where $\rho$-like and $\sigma$-like combinations of $\pi\pi$ can appear at relatively low energies.

The system we study is closely related to $S$-wave pseudoscalar-pseudoscalar scattering of $D\pi$, $D\eta$, $D_s\bar{K}$, considered in ref.~\cite{Moir:2016srx}. An added complication is that in the pseudoscalar-vector case partial waves of differing $\ell$ but the same $J$ can mix (also referred to as dynamical mixing) due to multiple possibilities to couple spin and orbital angular-momentum. Since parity is conserved and proportional to $(-1)^\ell$, mixing partial-waves always have a gap $\Delta \ell = 2$. We will thus simultaneously consider $\SLJ{2S+1}{\ell}{J}=\SLJ{3}{S}{1},\SLJ{3}{D}{1}$ scattering in $\dstarpi$ (the D wave is strongly threshold-suppressed in the heavier channels). We also consider $J^P=2^+$, which features $D\pi\SLJc{1}{D}{2}$ coupled to $D^\ast\pi\SLJc{3}{D}{2}$. In this channel we neglect the $D\eta$ and $D_s \bar{K}$ contributions as they could only appear in D wave, which is strongly suppressed in this region of energy which is near the respective thresholds. Negative-parity channels will be considered as well since they are needed to correctly analyse the finite-volume spectra with non-zero total momentum. The partial waves relevant to vector-pseudoscalar and pseudoscalar-pseudoscalar scattering that appear for a given $J^P$ combination are listed in table~\ref{tab:jp_pw}.

\begin{table}[ht]
	\begin{center}
			\begin{tabular}{l|l|l}	
				$J^P$  & $\SLJ{1}{\ell}{J}$ & $\SLJ{3}{\ell}{J}$\\
				\hline
				\hline
				$0^+$ & $\osz$ & \\
				$0^-$ &   & $\tpz$ \\
				$1^+$ &   & $\tso, \tdo$ \\
				$1^-$ &  $\opo$ & $\tpo$ \\
				$2^+$ &  $\odt$ & $\tdt$ \\
				$2^-$ &   & $\tpt, \SLJ{3}{F}{2}$ \\
				$3^+$ &   & $\tdthr, \SLJ{3}{G}{3}$\\
				$3^-$ & $\SLJ{1}{F}{3}$ & $\SLJ{3}{F}{3}$ \\[0.5ex]
			\end{tabular}
		\caption{Total angular momenta up to $J=3$ together with the corresponding partial waves
			for the case of pseudoscalar-pseudoscalar and vector-pseudoscalar scattering.}
		\label{tab:jp_pw}
	\end{center}
\end{table}

\subsection{Spectra from finite-volume correlation functions}

In lattice QCD, observables are obtained from a euclidean path integral using a discretized form of the QCD action. The fermionic part is evaluated by Grassmann integration as in the continuum theory, leaving an integral only over the gauge fields, which can be evaluated numerically through Monte-Carlo sampling of gauge field configurations. The discretization of spacetime regularizes the theory, with the finite lattice spacing acting as the ultraviolet cut-off. In this work, calculations are performed in a finite spatial volume $L^3$ with periodic boundary conditions, leading to a quantization of momenta and a discrete spectrum. The presence of interactions between hadrons in the finite volume causes shifts or additional levels in the quantized spectrum, and through the L\"uscher method~\cite{Luscher:1986pf} this can be used to determine the physical scattering amplitudes.

The finite-volume spectra are computed from two-point correlation functions of operators interpolating states that carry the quantum numbers of interest. There is significant freedom in the construction of the operators, which can be exploited to more reliably determine excited states. For a given set of interpolating operators (interpolators) $\set{\mathcal{O}_i}$ we compute a matrix of correlation functions
\begin{equation}
	C_{ij}(t) = \braket{0 | \mathcal{O}_i(t) \mathcal{O}_j^\dagger(0) | 0}\;.
\end{equation}
Solving the generalized eigenvalue problem (GEVP) $C_{ij}(t) v_j^{\mathfrak{n}} = \lambda^{\mathfrak{n}} (t,t_0) C_{ij}(t_0) v_j^{\mathfrak{n}}$ for a suitably chosen value of $t_0$ we obtain an eigenvector $v^\mathfrak{n}$ that defines a variationally optimised interpolator $\Omega_\mathfrak{n}^\dagger\sim\sum_i v^\mathfrak{n}_i\mathcal{O}_i^\dagger$ for state $\mathfrak{n}$~\cite{Michael:1985ne,Luscher:1990ck,Blossier:2009kd,Dudek:2009qf}. The time dependence of the eigenvalues, referred to as \emph{principal correlators}, exposes the spectrum when fitted by an exponential. We typically use the form
$\lambda^{\mathfrak{n}} (t,t_0) = (1-A_\mathfrak{n}) e^{-E_\mathfrak{n}(t-t_0)} + A_\mathfrak{n} e^{-E_\mathfrak{n}' (t-t_0)}$, where the second term absorbs remnant excited state contamination.

Correlation functions are computed using the distillation framework~\cite{Peardon:2009gh}, wherein the quark fields appearing in the operator constructions are smeared with the distillation operator, given by $\square(t) = \sum_{i=1}^{N_{\text{vec}}} v_i(t) v_i^\dagger(t)$, with $v_i$ representing the first $N_{\text{vec}}$ eigenvectors (ordered by eigenvalue) of the gauge-covariant Laplacian, evaluated on the gauge-field background. A good choice of $N_{\text{vec}}$ depends on the volume and determines the degree of smearing applied to the fields (see table~\ref{tab:ensembles} for the numbers of vectors used in this calculation). The distillation framework enables the efficient computation of all Wick contractions required by QCD, including diagrams with annihilation lines. In particular, propagators of distilled fields can be computed once and reused for a wide range of operator constructions, making this framework well suited for the computation of large correlation matrices to be used in the GEVP.

\subsection{Lattices}
\label{subsec:lattice_tec}

This calculation is performed on three ensembles of gauge configurations corresponding to three different spatial volumes (see table~\ref{tab:ensembles}). These ensembles have been generated using an anisotropic action, with the temporal lattice spacing being finer than the spatial one. This is a reasonable choice when computing spectral quantities as the finer temporal resolution allows more precise energy measurements.
An $\mathcal{O}(a)$-improved Wilson-clover fermion action is used with $2 + 1$ dynamical quark flavours, corresponding to two degenerate light-quarks and a heavier strange quark, which is tuned approximately to its physical value. Due to the degenerate up and down quark masses isospin is an exact symmetry on these lattices. In the gauge sector we use a tree-level Symanzik-improved action. The necessary tuning of the Wilson coefficients appearing in the improved actions is described in ref.~\cite{Edwards:2008ja,Lin:2008pr}. %
Periodic spatial and anti-periodic temporal boundary conditions are applied. We choose to set the scale by comparing the measurement of the $\Omega$-baryon mass from these ensembles to its physical value, leading to $a_t^{-1} = \tfrac{m_{\Omega}^{\text{phys.}}}{a_t m_{\Omega}^{\text{lat.}}} = 5667$~MeV~\cite{Edwards:2012fx}. The quantity used to set the scale is not unique and other methods may lead to slightly different values of $a_t^{-1}$.
The spatial lattice spacing is given by $a_s = \xi \cdot a_t \approx 0.12$~fm, where $\xi \approx 3.5$ quantifies the anisotropy. 
The valence charm-quarks use the same Wilson-clover action as the light and strange quarks with a mass parameter tuned to approximately reproduce that of the physical $\eta_c$ meson~\cite{Liu:2012ze}.

\begin{table}
	\centering
\begin{tabular}{c|c|c|c|c|c}
	$L/a_s$ & $T/a_t$  & $V_3$  & $N_{\text{cfg}}$ & $N_{\text{vec}}$ & $N_{\text{t-src}}$  \\
	\hline
	\hline
	16 & 128 & $(1.9\text{fm})^3$ & 478 & 64 & 4 \\
	20 & 128 & $(2.4\text{fm})^3$ & 603 & 128 & 1-4 \\
	24 & 128 & $(2.9\text{fm})^3$ & 553 & 160 & 1-4 \\
\end{tabular}
\caption{Lattice ensembles used in the calculation of the spectra. The columns contain the spatial extent $L$ and temporal extent $T$ of the lattice in units of the spatial and temporal lattice spacing respectively, the physical volume $V_3$, the number of gauge configurations $N_{\text{cfg}}$, the number of distillation vectors $N_{\text{vec}}$ and the number of time sources $N_{\text{t-src}}$, which are averaged. On all ensembles $m_\pi = 391$~MeV. The scale setting used to convert to physical units is explained in section~\ref{subsec:lattice_tec}.}
\label{tab:ensembles}
\end{table}

\subsection{Operator basis}

Single-hadron-like operators are formed from quark-bilinears of the form $\bar{\psi}\Gamma D...D\psi$ using up to three gauge-covariant derivatives $D$ at-rest, and up to two derivatives for non-zero momentum. These are projected onto definite $J$, as described in refs.~\cite{Dudek:2009qf,Dudek:2010wm}.
Since the finite cubic boundary does not preserve the rotational symmetry of infinite continuous space, we project operators to irreducible representations (irreps) $\Lambda$ of the appropriate reduced symmetry group, which depends on the direction of the overall momentum. Only certain values of $J^P$ from the continuum will contribute to a given $\Lambda$ in the finite volume. However, since there is a finite number of finite-volume-group irreps but an infinite number of values for $J$, every $\Lambda$ has an infinite number of subductions. Fortunately it is only necessary to consider the first few $J$. This becomes clear in the amplitude analysis, where partial waves of higher orbital angular momenta $\ell$ are suppressed by a factor of $k^{2\ell}$ close to threshold (where $k$ is the centre-of-mass-frame momentum).
Table~\ref{table:pw_irreps_dstarpi} lists the irreps of $O_h$ and its relevant subgroups as well as the lowest continuum angular momenta that subduce to it~\footnote{As we are only dealing with mesons in this paper we do not consider the double-cover of these groups.}.
A single-meson-like operator with flavour $\pmb{F}$ subduced from a continuum $J$ to irrep $\Lambda$ is given by
\begin{equation}
	{\mathcal{O}^\dagger}^{\Lambda \mu; [J]}_{\pmb{F}\nu} (\vec{p}, t) = \sum_m \mathcal{S}^{J,m}_{\Lambda, \mu} {\mathcal{O}^\dagger}^{Jm}_{\pmb{F}\nu} (\vec{p}, t)\;.
\end{equation}
where $\mathcal{S}^{J,m}_{\Lambda, \mu}$ denotes the subduction coefficient~\cite{Thomas:2011rh}.
Two-hadron-like operators are formed from pairs of single-hadron-like operators such that the overall operator has definite flavour quantum numbers, momentum and transforms as an irrep of the corresponding finite-volume group. The single-hadron operators used in the construction typically correspond to variationally optimal interpolators obtained from the GEVP of a lattice study with the corresponding quantum numbers. Schematically we have,
\begin{align}
\Omega_{\mathfrak{nm}}^\dagger(\vec{P}) \sim \sum_{\vec{P}=\vec{p}_1+\vec{p}_2} (\mathrm{CGs})\; \Omega_\mathfrak{n}^\dagger(\vec{p}_1)\;\Omega_\mathfrak{m}^\dagger(\vec{p}_2)\;,
\end{align}
where $\mathfrak{n}$ and $\mathfrak{m}$ label single-meson states, $(\mathrm{CGs})$ is a placeholder for lattice Clebsch-Gordan coefficients coupling the finite-volume irreps of the operators, $\vec{p}_1$ and $\vec{p}_2$ are the individual momenta of the single-meson operators and $\vec{P}$ is the total momentum of the two-meson operator.
This is described in refs.~\cite{Dudek:2012gj,Woss:2020ayi}.

In this work, we construct single-meson-like charmed $I=1/2$ operators, that produce $D$-meson-like states. We also construct two-meson like operators resembling a $D^{(\ast)}$ or $D_s^{(\ast)}$ combined with a $\pi$, $\eta$, $\bar{K}$ or $\eta^\prime$ in the relevant $I=1/2$ combinations.
We use the non-interacting energies as a guide to what meson-meson combinations should be included. In the absence of interactions the finite-volume energy associated to the meson-meson operators is given by
 \begin{equation}
	E=\sqrt{m_1^2 + |\vec{p_1}|^2} + \sqrt{m_2^2 + |\vec{p_2}|^2} \;,
\end{equation}
where $m_i$ are the energies of the single-meson ground states. These energies are listed in the left panel of table~\ref{tab:dstarpi:had_masses}. The threshold energies, i.e. $\vec{p}_1 = \vec{p}_2 = 0$, corresponding to relevant meson-meson operators, are listed in the right panel. We include all meson-meson operators corresponding to energies up to $\dstarpi\pi$ threshold in the spectrum computation. The full list is given in table~\ref{tab:app:dstarpi:ops2} in the appendix.

\begin{table}[ht]
	\begin{center}
			\begin{tabular}{ccc|l}	
				$\vec{d}$ & $G$ & $\Lambda$  & $J^P_{([N])}$\\
				\hline
				\hline
				\multirow{7}{*}{$[000]$}  & \multirow{7}{*}{$O_h$} & 
				$A_2^+$  & $3^+$ \\
				& & $E^+$    & $2^+$ \\
				& & $T_1^+$    & $1^+$, $3^+$  \\
				& & $T_2^+$    & $2^+$, $3^+$ \\
				& & $E^-$    & $2^-$ \\
				& & $T_1^-$  & $1^-$, $3^-$  \\
				& & $T_2^-$  & $2^-$, $3^-$ \\[0.5ex]
				\hline
				\multirow{3}{*}{$[n00]$}  & \multirow{3}{*}{$C_{4v}$} &
				$A_2$  & $0^-$, $1^+$, $2^-$, $3^+$  \\
				& &$B_1,\:B_2$  & $2^+$, $2^-$, $3^+$, $3^-$  \\
				& &$E_2$  & $1^+$, $1^-$, $2^+$, $2^-$, $3^+_{[2]}$, $3^-_{[2]}$ \\[0.5ex]
				\hline
				\multirow{2}{*}{$[nn0]$}  & \multirow{2}{*}{$C_{2v}$} &
				$A_2$  & $0^-$, $1^+$, $2^+$, $2^-_{[2]}$, $3^+_{[2]}$, $3^-$ \\
				& &$B_1,\:B_2$  & $1^+$, $1^-$, $2^+$, $2^-$, $3^+_{[2]}$, $3^-_{[2]}$ \\[0.5ex]
				\hline
				\multirow{2}{*}{$[nnn]$}  & \multirow{2}{*}{$C_{3v}$} & 
				$A_2$  & $0^-$, $1^+$, $2^-$, $3^+_{[2]}$, $3^-$  \\
				& &$E_2$  & $1^+$, $1^-$, $2^+_{[2]}$, $2^-_{[2]}$, $3^+_{[2]}$, $3^-_{[2]}$ \\[0.5ex]
			\end{tabular}
		\caption{Subgroups $G$ of $O_h$, corresponding to the residual rotation and reflection symmetry in a cubic volume moving in the direction $\vec{d}$, corresponding to an overall momentum $\vec{P} = \tfrac{2\pi}{L}\vec{d}$ of the system. Relevant irreps $\Lambda$ of the finite groups and the continuum angular momenta up to $J=3$ subducing in these irreps are listed. In some irreps there are $J^P$ combinations with more than one \emph{embedding} $N$, as indicated in the subscript.}
		\label{table:pw_irreps_dstarpi}
	\end{center}
\end{table}

 \begin{table}[htb!]
	\begin{center}
		\begin{minipage}[b]{0.48\textwidth}
			\begin{tabular}{c|c|c}
				& {$a_t m$} & $m$/MeV \\
				\hline
				\hline
				$\pi$       & 0.06906(13) &  391~\cite{Dudek:2012gj}  \\
				$K$         & 0.09698(9)  & 549~\cite{Wilson:2014cna}   \\
				$\eta$      & 0.10364(19) & 587~\cite{Dudek:2016cru}   \\
				$\eta^\prime$      & 0.1641(10)  & 930~\cite{Dudek:2016cru}   \\
				$D$         & 0.33303(31) &  1887~\cite{Cheung:2020mql} \\
				$D_s$       & 0.34441(29) &  1952~\cite{Cheung:2020mql}  \\
				$D^\ast$    & 0.35494(46) &  2011~\cite{Cheung:2020mql} \\
				$D_s^\ast$    & 0.36587(35) & 2073~\cite{Cheung:2020mql}  \\
			\end{tabular} 
		\end{minipage}  
		\hspace{0.02\textwidth}
		\begin{minipage}[b]{0.48\textwidth}
			\begin{tabular}{c|c|c}       
				& {$a_t E_\mathrm{threshold}$} & {$E_\mathrm{threshold}$/MeV}\\
				\hline
				\hline
				$D\pi$              & 0.4021(3) & $2279$ \\
				$D^\ast \pi$	    & 0.4240(5) & $2403$ \\
				$D\eta$              & 0.4367(4) & $2475$ \\
				$D_s \bar{K}$              & 0.4414(3) & $2501$ \\
				$D^\ast \eta$	    & 0.4586(5) & $2599$ \\
				$D_s^\ast \bar{K}$  & 0.4629(4) & $2623$\\
				$D\pi\pi$           & 0.4711(4) & $2670$\\
				$D^\ast \pi\pi$     & 0.4931(5) & $2794$ \\
			\end{tabular} 
		\end{minipage}
	\end{center}
	\caption{Relevant stable-meson masses (left-hand side) and two-particle thresholds (right-hand side) in lattice and physical units, as determined on the ensembles used in this calculation.}
	\label{tab:dstarpi:had_masses}
\end{table}

The single-meson energies have been extracted in previous lattice studies using a fit of the relativistic dispersion relation
\begin{equation}
	(a_t E)^2 = (a_t m)^2 + {|\vec{d}|}^2\left(\frac{2\pi}{\xi \;L/a_s}\right)^2
\end{equation}
to the ground state energies of the relevant meson determined at each momentum. The fits yielding the heavy meson masses used in this analysis are shown in figure 2 of ref.~\cite{Cheung:2020mql}. Apart from the mass $m$, the renormalized anisotropy $\xi$ is also a free parameter in this fit. It is needed when converting energies measured at non-zero overall momentum to the corresponding energies in the rest frame. We take the same approach outlined in ref.~\cite{Cheung:2020mql} using $\xi_\pi$ for the computation of our nominal spectra and the envelope around $\xi_\pi$, $\xi_D$ and $\xi_{D^\ast}$ including their fit uncertainties when estimating the systematic variations in the determination of the amplitudes in sec.~\ref{sec:scattering}.
Also following ref.~\cite{Cheung:2020mql}, we assign an additional uncertainty to all finite volume energy levels based on the observation that the extracted masses of the $D$ and $D^\ast$ mesons show some volume dependence beyond their fit uncertainties. From the envelope around $m_D$ and $m_{D^\ast}$ in the $20^3$ and $24^3$ volumes, we determine $a_t \delta E_{\text{syst.}} = 0.00030$, which is added to the statistical uncertainty on each level in quadrature.

\subsection{Finite volume spectra}
\label{subsec:fv_spectra}

\begin{figure}[htb!]
	\centering
	\includegraphics[width=\textwidth]{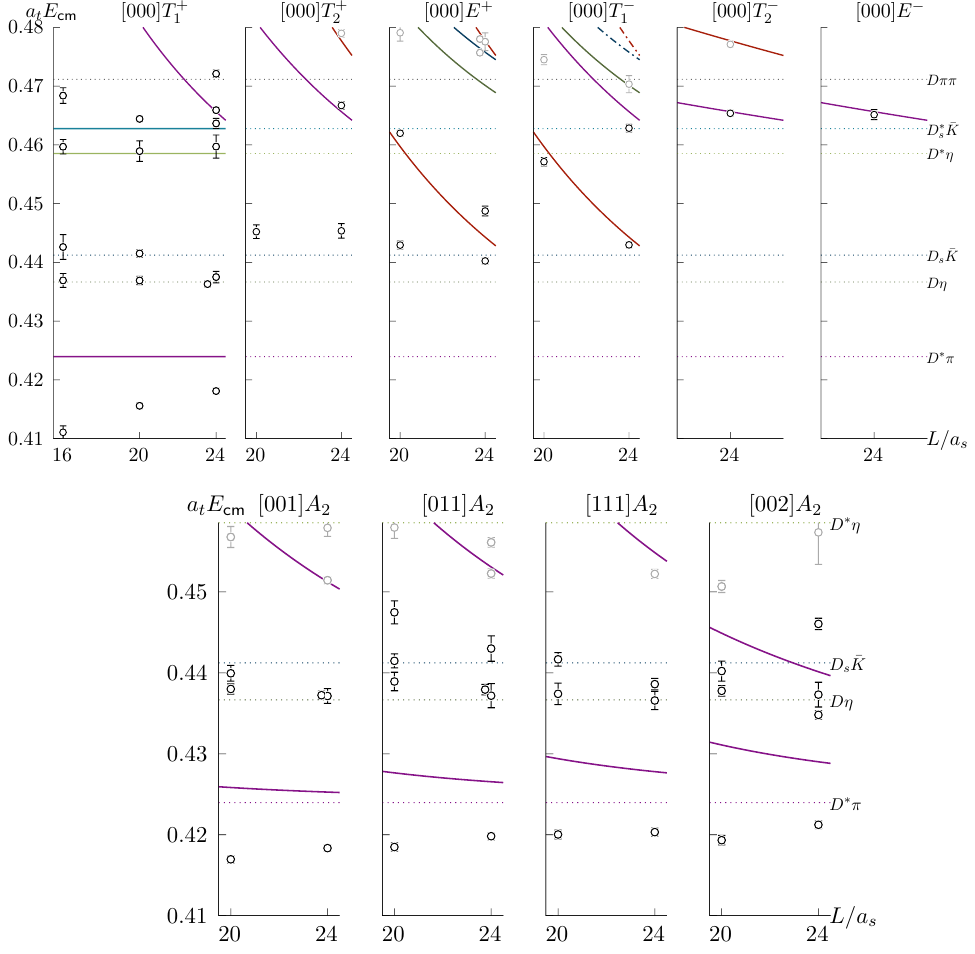}
	\caption[Finite-volume spectra]{Energy levels (circular markers with error bars) measured in finite spatial volumes of size $V_3/a_s^3 = 16^3$ (at rest only),  $20^3$ and $24^3$. Levels printed in grey are not considered in the amplitude fits of section~\ref{sec:scattering}. Non-interacting energies of meson-meson operators included in the calculation are given by solid coloured lines.  Two-meson thresholds are represented by dotted lines and operators that were excluded due to non-interacting energies beyond our cut-off by dashed-dotted lines. The moving-frame irreps have $J^P=0^-$ subductions.}
\label{fig:dstarpi:spec_i}
\end{figure}

\begin{figure}[htb!]
	\centering
	\includegraphics[width=\textwidth]{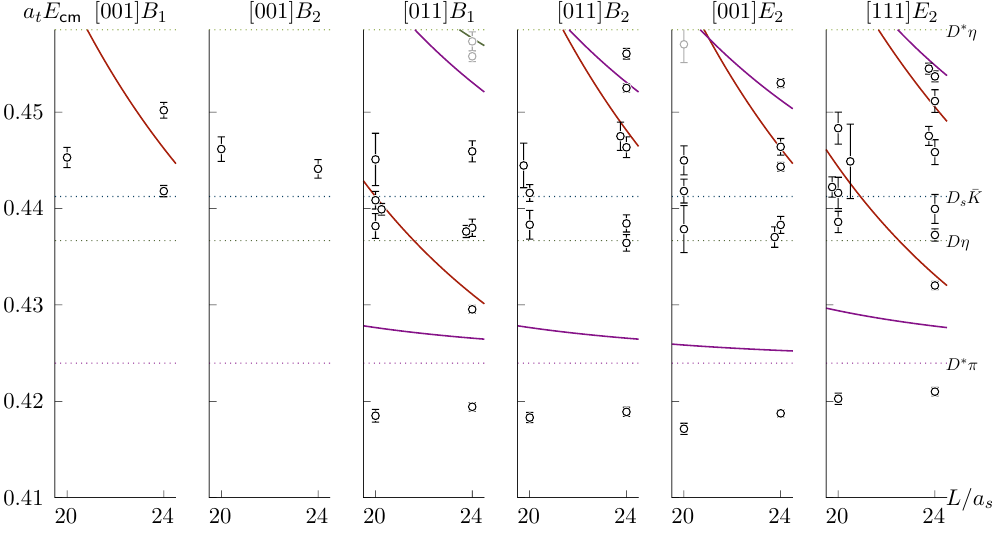}
	\caption[Moving-frame finite-volume spectra without $J^P=0^-$ subductions]{As figure~\ref{fig:dstarpi:spec_i} but for moving-frame irreps without $J^P=0^-$ subductions.}
\label{fig:dstarpi:spec_ii}
\end{figure}

We extract finite volume spectra with zero total momentum up to at least $a_t E_{\cm} = 0.48$, which is slightly above $D\pi\pi$ threshold ($E_{D\pi\pi}|_\thr$), and moving-frame spectra up to at least $E_{\dstareta}|_\thr$.  Not all computed levels are used in the scattering analysis. Figures~\ref{fig:dstarpi:spec_i} and \ref{fig:dstarpi:spec_ii} contain the spectra $\set{E_{\mathfrak{n}}}$ obtained from the variational analysis. The spectra are labelled according to the irreps listed in table~\ref{table:pw_irreps_dstarpi} and the magnitude of the overall momentum, which identifies the group, i.e. $[\vec{d}] \Lambda^{(P)}$. Parity $P$ is only indicated in irreps with zero total momentum ($O_h$), where it is a conserved quantum number. Together with isospin, strangeness and charm, this specifies a full set of quantum numbers in the finite volume. A quantitative analysis of the spectra 
will be performed in the next section, constraining the scattering amplitudes according to the interactions observed in the finite volume. However, some qualitative statements can already be made based on the partial-wave content of the irreps, the knowledge of the non-interacting energies and the counting of energy levels.

As explained above, each irrep contains an infinite tower of continuum angular momentum subductions. However, the angular-momentum barrier leads to a suppression of partial-wave amplitudes close to threshold in the absence of resonant enhancements. It is then sufficient to consider only the lowest few partial waves in a given irrep. Generally, if we find the effect of partial-wave $\ell$ to be small we argue that partial-wave $\ell+1$ may be neglected unless the channel has a nearby bound state or resonance.

For the axial-vector channel, the low-lying $[000]T_1^+$ spectrum is the easiest to interpret. The relevant $\dstarpi$ partial-waves are the dynamically-mixing pair $\set{\tso, \tdo}$ and $\tdthr$. The latter can also be determined from the $[000] A_2^+$ spectrum, where the lowest energy level in the $L=24$ volume is found at $a_t E_\cm \approx 0.498$, far beyond the energy range that we study, and coincides with the non-interacting curve of the corresponding $\dstarpi$ operator, indicating insignificant $\tdthr$ interactions.
Below $E_{\dstareta}|_\thr$ there are two additional levels in $[000]T_1^+$ with respect to the expectation in absence of interactions. The lowest energy levels are shifted down with respect to the $\dstarpi$ threshold. This is a strong indication for the presence of bound states and/or resonances in this channel.
The tensor states subduce in $[000]E^+$ and $[000]T_2^+$, where we also find an additional finite-volume level.
These patterns repeat in the irreps of the non-zero momentum subgroups, in the sense that irreps which have a contribution by the $J^P=1^+$ channel exhibit two extra levels, those with $J^P=2^+$ have one extra level and those which contain subductions of both have three extra levels. In the case of $[111]E_2$, $J^P=2^+$ appears \emph{twice}. Beyond level-counting, the mixing of more partial waves in these irreps makes a qualitative interpretation rather difficult.

Among the negative-parity irreps, $[000]T_1^-$, where the mixing vector-like $\dpi\{\opo\}$ and $\dstarpi\{\tpo\}$ partial waves are the dominant contribution, exhibits the largest shifts in energies albeit no extra levels between $E_{\dpi}|_\thr$ and $E_{\dpi \pi}|_\thr$. Note that there is a level below the kinematic threshold of $\dpi$ scattering at $a_t E_{\cm} \approx 0.355$ in all irreps that feature a $J^P=1^-$ subduction. This level can be associated with the vector $D^\ast$ bound state~\cite{Moir:2016srx} and we will ignore it in the subsequent analysis, as it is too far below threshold to significantly affect $\dstarpi$ scattering. The levels in  $[000] T_2^-$ and $[000] E^-$ are compatible with the non-interacting spectrum and suggest little to no interactions in $J^P = 2^-$ which corresponds to $\tpt$ in the $\dstarpi$ channel.


%% file: scattering.tex
%
\section{Scattering amplitudes}
\label{sec:scattering}

We now use the energies presented in section~\ref{sec:lattice} to constrain $I=1/2$, $C=1, S=0$ scattering amplitudes. In our previous work~\cite{Lang:2022elg}, $D^\ast\pi$ in a dynamically-coupled $\SLJ{3}{S}{1}-\SLJ{3}{D}{1}$ system was determined by strictly considering only energy levels below $D^\ast\eta$ threshold. This work begins where that study ended by relaxing this restriction and considering higher energies up to the first three-hadron threshold, $D\pi\pi$.\footnote{The lower energy levels common to both this and the previous analysis in ref.~\cite{Lang:2022elg} are identical.} In doing so, $D^\ast\eta$ and $D_s^\ast\bar{K}$ amplitudes can also be constrained, which have similarities to $D\eta$ and $D_s\bar{K}$ that were determined on these ensembles previously~\cite{Moir:2016srx}. Including higher energy levels also improves the determination of higher partial waves, which benefit from a few relatively isolated extra levels obtained from irreps at-rest.
We first review the methods used and introduce the coupled-channel scattering amplitudes.

\subsection{Formalism}
For arbitrary 2-to-2 particle scattering processes in a finite volume the following quantisation condition (QC) holds up to corrections that are exponentially suppressed by the spatial extent $L$~\cite{Briceno:2014oea},
\begin{align}
	\det\Bigl[\bm{1}+i \bm{\rho}(E_\cm)\cdot\bm{t}(E_\cm)\cdot\bigl(\bm{1}+i\bm{\mathcal{M}}(E_\cm,L)\bigr)\Bigr]=0 \;.
	\label{eq_det}
\end{align}
All quantities are matrices in the space of open scattering channels and partial waves. $\bm{\mathcal{M}}(E_\cm,L)$ is a matrix of functions of the $\cm$-frame energy and the spatial extent $L$ of the finite volume which is completely determined by the masses and spins of the scattered particles. It is a finite-volume quantity and therefore in general not diagonal in angular momentum. $\bm{\rho}(E_\cm)$ is a diagonal matrix of relativistic two-body phase space factors with entries given by $\rho_{ij} = \frac{2k_i}{E_\cm} \delta_{ij}$, where $i$ and $j$ label open channels. $k_i$ is the $\cm$-frame momentum in scattering channel $i$,
\begin{equation}
		k_i^2(E_\cm; m_i^{(1)}, m_i^{(2)}) = \frac{1}{4E_\cm^2} \left(E_\cm^2 - (m_i^{(1)} + m_i^{(2)})^2\right) \left(E_\cm^2 - (m_i^{(1)} - m_i^{(2)})^2\right) \;,
\end{equation}
where $m_i^{(1)}$ and $m_i^{(2)}$ are the masses of the scattered hadrons. The quantity of interest is $\bm{t}(E_\cm)$, the infinite-volume scattering matrix, which is block diagonal in total angular momentum $J^P$, but can couple hadron-hadron scattering channels and/or partial wave amplitudes with the same $J^P$ but different $\ell$.

In the special case of elastic scattering and no mixing between partial waves, the $t$ matrix can be parameterised by a scattering phase shift $\delta_\ell(E_\cm)$ for each partial wave $\ell$,
\begin{equation}
	t_{\ell} = \frac{1}{\rho} e^{i \delta_\ell} \sin \delta_\ell \;.
\end{equation}
If only a single partial wave is relevant for the scattering process the phase shift at a given finite-volume energy can be directly obtained from the quantisation condition, which simplifies to
\begin{equation}
		\cot \delta(E) = \mathcal{M}(E,L) \;. 
		\label{eq:elastic_qc}
\end{equation}
In a multi-channel scattering problem, or taking into account several partial waves that mix in the finite volume, the QC is underdetermined for a single energy level, making it necessary to parametrise the $t$ matrix as a function of energy.
A flexible way to do this, allowing for arbitrarily many channels and possibly overlapping resonances, is the $K$-matrix formalism.
The inverse $\bm t$ matrix is written
\begin{align}
	[{\bm t}^{-1}]_{ij} = (2k_i)^{-\ell_i}\: [{\bm K}]_{ij}^{-1}\:(2k_j)^{-\ell_j} + {\bm I}_{ij}\;,
	\label{eq_tmat}
\end{align}
where $\bm{K}$ is a real symmetric matrix with no branch cuts as a function of energy and $\bm{I}$ is a diagonal matrix with $\text{Im} I_{ij} = -\rho_i \delta_{ij}$ above $E_i|_\thr$.
The $S$ matrix in this formalism is manifestly unitary.
One can choose $\text{Re} I_{ij} = 0$. Improved analytic properties of the amplitude below threshold and at complex values of the energy (see section~\ref{sec:poles}) can be achieved by using the Chew-Mandelstam (CM) phase space~\cite{Wilson:2014cna}. In this case the real part of $\bm{I}$ is obtained from the imaginary part as the solution of a dispersion integral. The integral is generally divergent and needs to be regularised by a subtraction at some arbitrary energy. Our amplitudes in this section and the next use the CM phase space and we subtract either at the energy of the relevant threshold for channel $i$, or the value of the lowest pole mass parameter, if present.
The factors of $(2k_i)^{-\ell_i}$ in equation~\ref{eq_tmat} account for the expected threshold behaviour of the amplitude considering the angular momentum barrier.
The QC is used to predict the spectrum in a finite-volume for a given $t$-matrix parametrisation and corresponding set of parameter values. In practice we use an eigenvalue decomposition of the matrix in eq.~\ref{eq_det} in order to determine the spectrum from a given $\bm t$. Details can be found in ref.~\cite{Woss:2020cmp}. The parameters can be adjusted using a suitable fitting algorithm such that the predicted spectrum matches the spectrum found in the lattice calculation. In this analysis we use a $\chi^2$-minimisation procedure based on \emph{Minuit}.  The parameters corresponding to the $\chi^2$ minimum approximate the amplitude in the infinite volume~\cite{Guo:2012hv,Dudek:2012xn,Dudek:2014qha,Wilson:2014cna,Briceno:2017max}.

\subsection{Coupled-channel axial-vector $S$ and $D$ waves from $[000]T_1^+$}
\label{sec:coupled_1p}

The $J^P = 1^+$ amplitude was determined for a single hadron-hadron channel in ref.~\cite{Lang:2022elg}. Here, we extend the $K$-matrix definition to include the $\dstareta$ and $\dsstarkbar$ channels. This amplitude may be constrained using the spectrum of the zero-momentum $T_1^+$ irrep alone. Even though our final result will be based on a fit to the full spectrum, this is a valuable cross-check. The mixing between partial waves in the finite volume can lead to ambiguities in the fit, in particular when matching lattice energy levels with those determined by the QC. Finite-volume irreps at zero-momentum have fewer relevant contributing partial-waves which simplifies the fit.

\begin{figure}[htb!]
	\centering
	\includegraphics[width=\textwidth]{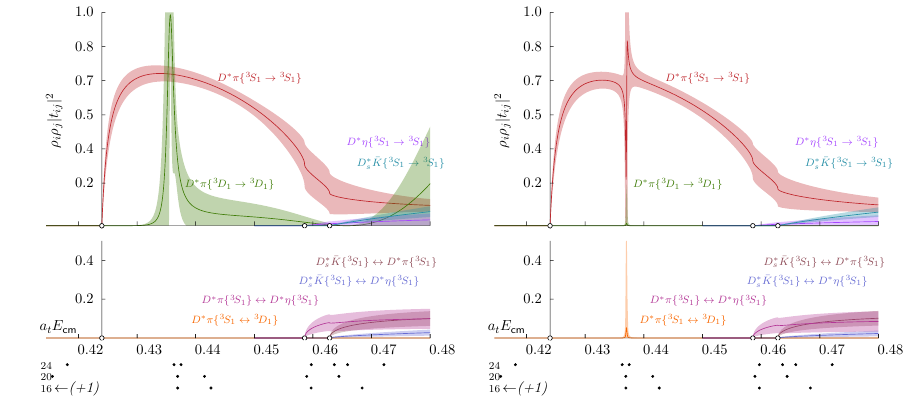}
	\caption{Squared amplitudes with angular momentum and parity $J^P = 1^+$. Contributing channels are $\dstarpi$, $\dstareta$ and $\dsstarkbar$ in S-wave and $\dstarpi$ in D wave. The upper panels contain the elastic parts of the amplitude and the lower panels the mixing. Error bands represent fit uncertainties. The energy levels from $[000] T_1^+$ constraining the amplitudes are displayed below the horizontal axis. The left and right-hand panels correspond to two distinct solutions of the QC that reproduce the spectrum in $T_1^+$, given by the parameters in equation~\ref{eq:1p_fit_a} and~\ref{eq:1p_fit_b} respectively.}
	\label{fig:dstarpi:amp_elastic_T1p}
\end{figure}

The lowest two angular-momentum subductions in $[000]T_1^+$ are $J^P = 1^+$ and $3^+$. $\dstarpi$ can form a $3^+$ either as a $\tdthr$ or a $\tgt$ configuration, the latter being suppressed by a factor of $k^4$ with respect to $D$ wave. $\tdthr$ can be estimated from the lowest level in the $L/a_s = 24$ spectrum of $[000] A_2^+$, which, using the elastic QC (equation~\ref{eq:elastic_qc}), gives a phase shift consistent with zero,
\begin{equation*}
	\delta_{\dstarpi\{\tdthr\}} (a_t E_\cm \approx 0.497) = (2.3 \pm 2.3)^\circ \;.
\end{equation*}
We conclude that $\dstarpi\{\tdthr\}$ (and any higher partial waves) can be neglected below this energy.
This suggests that the finite-volume spectrum in $[000] T_1^+$, consisting of 17 energy levels below our cut-off at $a_t E_{\cm} = 0.473$, may be well described by a $K$-matrix featuring the two mixing $\dstarpi \{\tso\}$ and $\dstarpi \{\tdo\}$ amplitudes and the heavier meson-meson channels $\dstareta$ and $\dsstarkbar$, for which we only include S wave (D wave is strongly suppressed at these energies). Given the two extra levels in the spectrum (with respect to the non-interacting energies), a suitable parametrisation is expected to be of the form
\begin{equation}
	\begin{aligned}
		&{\bm K}_{J^P = 1^+} = \sum_{r \in \set{0,1}} \frac{1}{m_r^2-s} \times\\ 
	&\begin{bmatrix}
		g^2_{r,\dstarpi \{\tso\}} & g_{r,\dstarpi \{\tso\}} g_{r,\dstarpi\{\tdo\}} & g_{r,\dstarpi \{\tso\}} g_{r,\dstareta \{\tso\}}  & g_{r,\dstarpi \{\tso\}} g_{r,\dsstarkbar \{\tso\}} \\
		& g^2_{r,\dstarpi \{\tdo\}} &  g_{r,\dstarpi\{\tdo\}} g_{r,\dstareta \{\tso\}} &  g_{r,\dstarpi\{\tdo\}} g_{r,\dsstarkbar \{\tso\}}  \\
		& & g^2_{r,\dstareta \{\tso\}} & g_{r,\dstareta \{\tso\}} g_{r,\dsstarkbar \{\tso\}} \\
		& & & g^2_{r,\dsstarkbar \{\tso\}}
	\end{bmatrix}\\
	& + \begin{bmatrix}
		\gamma_{\dstarpi \{\tso\}} & \gamma_{\dstarpi \{\tso \leftrightarrow \tdo\}} & \gamma_{\dstarpi \{\tso\} \leftrightarrow \dstareta \{\tso\}}  & \gamma_{\dstarpi \{\tso\} \leftrightarrow \dsstarkbar \{\tso\}} \\
		& \gamma_{\dstarpi \{\tdo\}} &  \gamma_{\dstarpi \{\tdo\}\leftrightarrow \dstareta \{\tso\}} &  \gamma_{\dstarpi \{\tdo\} \leftrightarrow \dsstarkbar \{\tso\}} \\
		& & \gamma_{\dstareta \{\tso\} } & \gamma_{\dstareta \{\tso\} \leftrightarrow \dsstarkbar \{\tso\}} \\
		& & & \gamma_{\dsstarkbar \{\tso\}}
	\end{bmatrix}\;.
	\end{aligned}
	\label{eq:kmat_coupled}
\end{equation}
The index $r$ labels pole terms. The mass of pole $r$ is parametrised by $m_r$, whereas $g_{r,i}$ introduces a coupling of this pole to channel/partial-wave $i$. Clearly, if two channels couple to the same pole the corresponding amplitudes will mix in the vicinity of the pole. The $\gamma_{a_i \{\SLJ{3}{(\ell_i)}{1}\}}$ describe the smooth part of the $K$-matrix element for the meson-meson combination $a_i$ and partial-wave $\ell_i$. The $\gamma_{a_i\{\SLJ{3}{(\ell_i)}{1}\}\leftrightarrow a_j\{\SLJ{3}{(\ell_j)}{1}\}}$ (or $\gamma_{a_i \{\SLJ{3}{(\ell_i)}{1}\leftrightarrow \SLJ{3}{(\ell_j)}{1}\}}$) are used to explicitly allow for mixing between channels $i$ and $j$ (mixing between partial-waves $\ell_i$ and $\ell_j$ for the same meson-meson combination $a_i$, respectively).
Typically only a subset of these parameters are significantly non-zero and with a small dataset not all of them can be constrained simultaneously. In this fit we introduce $\gamma$-terms only for the elastic part of the $\dstarpi$ and $\dsstarkbar$ $S$-wave amplitudes and fix the couplings of the two higher channels to the second pole to zero. These restrictions will be relaxed later.
Two distinct solutions can be obtained with implications for the shape of the $\dstarpi\{\SLJ{3}{D}{1}\}$ amplitude and its mixing with $\dstarpi\{\SLJ{3}{S}{1}\}$.
Fixing $g_{1,\dstarpi \{\tso\}} = 0$ but allowing for a non-zero $\gamma_{\dstarpi\{\tdo\}}$ the following parameters are determined in the $\chi^2$ minimization,
\begin{equation*}
	\resizebox{\textwidth}{!}{
		$\begin{aligned}[t]
			\begin{array}{rcl}
				g_{0,{D_s^*} \bar{K} \{\tso\}} & = & (-0.341 \pm 0.075) \cdot a_t^{-1}\\
				g_{0,\dstareta \{\tso\}} & = & (-0.25 \pm 0.11) \cdot a_t^{-1}\\
				g_{1,\dstarpi \{\tdo\}} & = & (10.2 \pm 4.4) \cdot a_t\\
				g_{0,\dstarpi \{\tso\}} & = & (0.550 \pm 0.069) \cdot a_t^{-1}\\
				\gamma_{{D_s^*} \bar{K}\{\tso\}} & = & (-0.92 \pm 0.58) \\
				\gamma_{\dstarpi\{\tdo\}} & = & (3800 \pm 3200) \cdot a_t^4\\
				\gamma_{\dstarpi\{\tso\}} & = & (2.40 \pm 0.72) \\
				m_0 & = & (0.42277 \pm 0.00034) \cdot a_t^{-1}\\
				m_1 & = & (0.43582 \pm 0.00073) \cdot a_t^{-1}\\
			\end{array}
			\end{aligned}
	\qquad
	\begin{aligned}[t]
		\begin{bmatrix}
			1.00 & 0.42 & -0.02 & -0.76 & -0.38 & -0.02 & -0.21 & -0.06 & 0.02\\
			& 1.00 & -0.13 & -0.78 & -0.16 & -0.12 & -0.01 & -0.20 & 0.06\\
			&  & 1.00 & 0.13 & -0.10 & 1.00 & 0.15 & 0.01 & -0.86\\
			&  &  & 1.00 & 0.21 & 0.13 & 0.46 & 0.02 & -0.13\\
			&  &  &  & 1.00 & -0.09 & -0.20 & -0.25 & -0.00\\
			&  &  &  &  & 1.00 & 0.17 & 0.00 & -0.86\\
			&  &  &  &  &  & 1.00 & -0.11 & -0.25\\
			&  &  &  &  &  &  & 1.00 & 0.15\\
			&  &  &  &  &  &  &  & 1.00\\
		\end{bmatrix}
	\end{aligned}$}
\end{equation*}
\begin{equation}
	\chi^2/N_{\text{dof}} = \tfrac{3.31}{17 - 9} =  0.41\;,
	\label{eq:1p_fit_a}
\end{equation}
where the uncertainties are statistical and the matrix on the right-hand side contains the correlation coefficients between the parameters.
The corresponding amplitude is shown in the left-hand panel of figure~\ref{fig:dstarpi:amp_elastic_T1p}.  If instead $g_{1,\dstarpi \{\tso\}}$ is allowed to float in the fit and $\gamma_{\dstarpi\{\tdo\}}$ fixed to zero the following set of parameters is obtained, the fit quality being identical:
\begin{equation*}
	\resizebox{\textwidth}{!}{
		$\begin{aligned}[t]
			\begin{array}{rcl}
				g_{0,{D_s^*} \bar{K} \{\tso\}} & = & (-0.334 \pm 0.075) \cdot a_t^{-1}\\
				g_{0,\dstareta \{\tso\}} & = & (-0.23 \pm 0.11) \cdot a_t^{-1}\\
				g_{1,\dstarpi \{\tdo\}} & = & (0.8 \pm 3.7) \cdot a_t \\
				g_{0,\dstarpi \{\tso\}} & = & (0.522 \pm 0.064) \cdot a_t^{-1}\\
				g_{1,\dstarpi \{\tso\}} & = & (-0.036 \pm 0.021) \cdot a_t^{-1}\\
				\gamma_{{D_s^*} \bar{K}\{\tso\}} & = & (-0.84 \pm 0.60) \\
				\gamma_{\dstarpi\{\tso\}} & = & (2.10 \pm 0.65) \\
				m_0 & = & (0.42268 \pm 0.00034) \cdot a_t^{-1}\\
				m_1 & = & (0.43715 \pm 0.00068) \cdot a_t^{-1}\\
			\end{array}
		\end{aligned}
		\qquad
		\begin{aligned}[t]
			\begin{bmatrix}
				1.00 & 0.39 & 0.10 & -0.76 & 0.10 & -0.49 & -0.18 & -0.05 & -0.02\\
				& 1.00 & -0.07 & -0.74 & 0.11 & -0.18 & 0.05 & -0.20 & -0.10\\
				&  & 1.00 & -0.11 & 0.01 & -0.18 & -0.32 & 0.22 & 0.07\\
				&  &  & 1.00 & -0.06 & 0.31 & 0.42 & 0.01 & -0.10\\
				&  &  &  & 1.00 & -0.00 & -0.11 & 0.04 & -0.60\\
				&  &  &  &  & 1.00 & -0.12 & -0.23 & -0.10\\
				&  &  &  &  &  & 1.00 & -0.14 & -0.21\\
				&  &  &  &  &  &  & 1.00 & 0.13\\
				&  &  &  &  &  &  &  & 1.00\\
			\end{bmatrix}
		\end{aligned}$}
\end{equation*}
\begin{equation}
	\chi^2/\Ndf = \tfrac{3.30}{17 - 9} =  0.41\;.
	\label{eq:1p_fit_b}
\end{equation}
The corresponding amplitude is shown in the right-hand panel of figure~\ref{fig:dstarpi:amp_elastic_T1p}. 
In both cases the $\dstarpi$ $S$-wave amplitude rises steeply from threshold and remains large throughout the elastic scattering region before dipping down close to the $\dstareta$ channel opening. The $\dsstarkbar$ $S$-wave is found to be small but non-zero whereas the $\dstareta$ $S$-wave is consistent with zero. Some amount of mixing between the three channels is found.

The $J^P=1^+$ $\tdo$ amplitude exhibits the shape of a narrow resonance. A non-zero width is found for this resonance when $g_{1,\dstarpi \{\tso\}} = 0$ and $\gamma_{\dstarpi\{\tdo\}}$ is allowed to float. In this case, the fit converges to a significantly non-zero value for $g_{1,\dstarpi \{\tdo\}}$. On the other hand, when $\gamma_{\dstarpi\{\tdo\}} = 0$ and $g_{1,\dstarpi \{\tso\}}$ is a free fit-parameter, the resulting $g_{1,\dstarpi \{\tdo\}}$ is consistent with zero and so is the width of the resonance. In this case there is significant mixing between $S$ and $D$ wave at the energy of the resonance.

One reason for these distinct solutions in $\tdo$ appears to be a single energy level constraining the $D$-wave tail to be zero just above the $\dsstarkbar$ channel-opening on the $L/a_s=24$ volume. This constraint can be satisfied either by a larger resonance width with a smooth background term, or by a near-zero coupling parameter leading to a vanishingly small resonance width.
This behaviour is in part due to this type of amplitude. In general, the resonance term in $\bm K$, in conjunction with the $D$-wave $k_i^{2\ell}$ threshold factors, do not produce an amplitude that goes to zero, even many resonance-widths away in energy. Barrier factors and higher order $K$-matrix polynomials can be used to remedy this, however these often produce spurious singularities. We will return to this topic later when considering a larger number of energy levels and variations of the parametrisation.
We investigate the sensitivity of these amplitudes to the specific choices for $\bm K$ later in section~\ref{sec:reference_amplitude}.

In order to improve these amplitude determinations, we make use of additional lattice irreps, some with non-zero total momentum. Due to the mixing of continuum $J^P$ in a finite cubic volume, it is necessary to also determine many other partial waves.

\subsection{Tensor channel from at-rest and non-zero-momentum spectra}
\label{sec:coupled_2p}

The tensor channel contains a single isolated $J^P=2^+$ resonance as was determined in ref.~\cite{Lang:2022elg}. For this study we add two energy levels not previously used that are relevant for this channel in the energy region above $D^\ast\eta$ threshold, one from $[000]T_2^+$ on $L/a_s=24$ close to a $D^\ast\pi$ non-interacting energy and another from $[000]E^+$ on $L/a_s=20$ close to a $D\pi$ non-interacting energy. There are no $D^{(\ast)}\eta$ or $D_s^{(\ast)}\bar{K}$ non-interacting levels in the energy region of interest and so these will not be constrained in $J^P=2^+$.\footnote{Given the selection of energy levels used, this will not affect the determination of the $1^+$ wave later.}

\begin{figure}[htb!]
	\centering
	\includegraphics[width=0.66\textwidth]{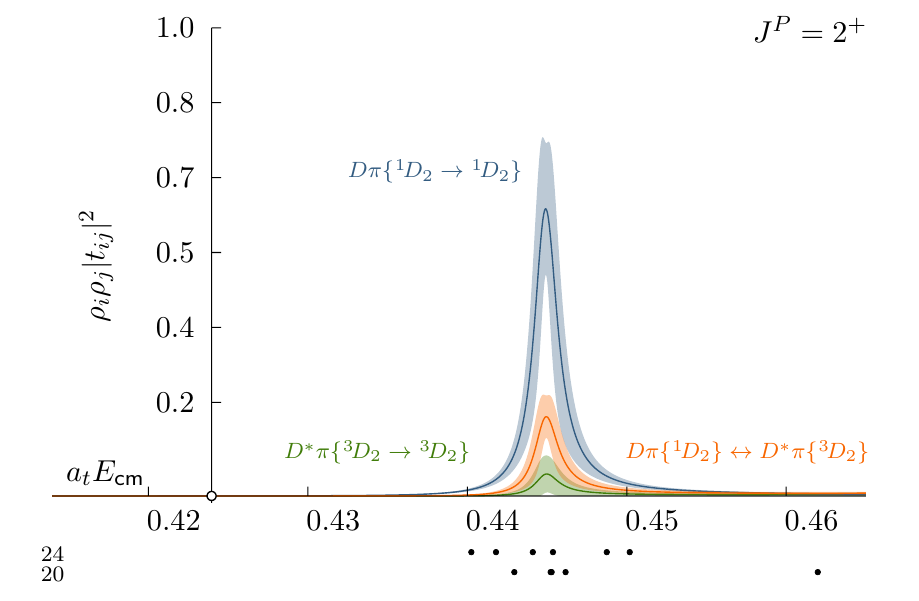}
	\caption{As figure~\ref{fig:dstarpi:amp_elastic_T1p} but showing $J^P = 2^+$ $\dstarpi\{\tdt\}$ and $D\pi \{\odt\}$ squared amplitudes constrained by energy levels from $[000] E^+$, $[000] T_2^+$, $[001] B_1$ and $[001] B_2$.}
	\label{fig:dstarpi:amp_extended_separate_2+}
\end{figure}

The leading contributions in the $J^P=2^+$ tensor channel are the coupled $\dpi\{\odt\}$ and $\dstarpi\{\tdt\}$ amplitudes. At zero momentum, $J^P=2^+$ subduces in the $[000] E^+$ and $[000] T_2^+$ irreps.  $[000] T_2^+$ also contains $J^P=3^+$, corresponding to $\dstarpi\{\tdthr\}$, which we found to be negligible. Any other partial waves in these irreps are expected to be further suppressed, and are not used. We also include the lowest levels from $[001] B_1$ and $[001] B_2$, where $2^+$ is present. $J^P=2^-$ from a $D^\ast\pi\SLJc{3}{P}{2}$ can also appear, however we will later find that this is also very small. The $J^P=2^+$ amplitudes can be described by a $K$ matrix of the two $D$-wave meson-meson channels. We use the form
\begin{equation}
	\begin{aligned}
		&{\bm K}_{J^P = 2^+} = \frac{1}{m^2-s}
		\begin{bmatrix}
			g^2_{\dpi \{\odt\}} & g_{\dpi \{\odt\}} g_{\dstarpi \{\tdt\}} \\
			& g^2_{\dstarpi \{\tdt\}}
		\end{bmatrix} +
		\begin{bmatrix}
			\gamma_{\dpi \{\odt\}} & \gamma_{\dpi \{\odt\} \leftrightarrow \dstarpi \{\tdt\}}  \\
			& \gamma_{\dstarpi \{\tdt\}}
		\end{bmatrix}\;.
	\end{aligned}
	\label{eq:kmat_2p}
\end{equation}
where the pole term is motivated by the observation of an extra level in any irrep containing $2^+$.
Here we fix $\gamma_{\dstarpi \{\tdt\}} = 0$ and $\gamma_{\dpi \{\odt\} \leftrightarrow \dstarpi \{\tdt\}} = 0$ but will vary this when including more energy levels later. The fit yields
\begin{equation*}
	\begin{aligned}[t]
		\begin{array}{rcl}
			g_{D \pi \{\odt\}} & = & (1.804 \pm 0.075) \cdot a_t\\
			g_{\dstarpi \{\tdt\}} & = & (2.54 \pm 0.64) \cdot a_t\\
			\gamma_{D \pi \{\odt\}} & = & (72 \pm 33) \cdot a_t^4 \\
			m & = & (0.44493 \pm 0.00044) \cdot a_t^{-1}\\
		\end{array}
	\end{aligned}
	\qquad
	\begin{aligned}[t]
		\begin{bmatrix}
			1.00 & 0.02 & 0.67 & 0.21\\
			& 1.00 & -0.04 & 0.03\\
			&  & 1.00 & -0.07\\
			&  &  & 1.00\\
		\end{bmatrix}
	\end{aligned}
\end{equation*}
\begin{equation}
	\chi^2/N_{\text{dof}} = \tfrac{6.31}{12 - 4} =  0.79\;.
\end{equation}
As can be seen from figure~\ref{fig:dstarpi:amp_extended_separate_2+} the amplitude has the form of an isolated resonance located at $a_t E_{\cm} \approx 0.445$. The $\dpi\{\odt\}$ amplitude is dominant over the $\dstarpi\{\tdt\}$, which may be attributable to the angular momentum barrier and the close proximity of the $\dstarpi$ threshold. This simple form naturally produces mixing between the channels when the $g_i$ parameters are nonzero. 

Further constraint can be obtained from other moving frame irreps where $J^P=2^+$ appears at the same time as $J^P=1^+$ as was done in ref.~\cite{Lang:2022elg}. We next consider the $J^-$ waves which we find to be non-resonant and relatively small in this energy region. It will then be possible to perform a combined determination of all the low partial waves.

\subsection{$P$-wave $J^-$ ``background'' amplitudes}
\label{sec:background}

The $J^P=0^-$, $1^-$ and $2^-$ channels appear as backgrounds in several of the moving-frame irreps. We estimate the importance of some of these contributions by looking at irreps at rest where they appear in isolation. The consistency between these estimates and the importance of the corresponding processes in the final fit is a good sanity check, as the finite-volume spectra in irreps at non-zero momentum constrain multiple partial waves and ambiguities can arise.

The $\dstarpi\{\tpz\}$ partial wave in $J^P=0^-$ is difficult to constrain in isolation. This is because the first energy levels in the at-rest irrep $[000] A_1^-$ where $0^-$ subduces are near the threshold for three-body $D\pi \pi$ scattering, which we cannot consider using the two-body quantisation condition. There are various sub-channel resonances in $D \pi \pi$ that make the $0^-$ spectrum rather complicated. On the one hand, the isoscalar $f_0$ could form a $0^-$ state with the $D$ meson in a relative $S$~wave, with no threshold suppression. However, this channel would be expected to open around
$a_t E_{\cm} \approx 0.546$ according to ref.~\cite{Briceno:2017qmb} which is far above the energy range of this analysis. More important is the $D_0^\ast$, which could form a $0^-$ state with the pion in an $S$~wave. This channel again has no threshold suppression and opens at
$a_t E_{\cm} \approx 0.47$ based on the determination of the $D_0^\ast$ at this pion mass in ref.~\cite{Moir:2016srx}. 
 
This means that care must be taken in using spectra from irreps that have a $0^-$ subduction. We will therefore refrain from using energies in the $[00n]A_2$, $[011]A_2$ and $[111]A_2$ irreps near the $D \pi \pi$ threshold and constrain the $1^+$ in the coupled-channel region using energy levels in $[000] T_1^+$ alone, where $0^-$ does not appear. Furthermore we simultaneously use energies from various moving-frame irreps with no $0^-$ subduction. The consistency of our solution across different spectra will then provide the necessary reassurance that $0^-$ scattering is a sub-leading process in the region of energy that we consider. Any residual effects can later be accounted for with a constant in the $K$ matrix parametrisation.

For $J^P = 1^-$ we can turn to the spectrum in $[000] T_1^-$ with three energy levels. 
This channel contains the mixing $\dstarpi\{\tpo\}$ and $\dpi \{\opo\}$ amplitudes.
Individual estimates for the elastic phase shifts can be obtained assuming that the mixing between the channels is small. For $L/a_s = 20$ the $[000] T_1^-$ irrep contains a level at $a_t E_{\cm} \approx 0.457$, slightly shifted down with respect to the nearby $\dpi$ non-interacting curve. The corresponding eigenvector from the GEVP suggests that this operator predominantly overlaps with $\dpi$ states.
This level is also near the upper limit of the elastic region in $J^P = 1^+$. Determining the $\dpi$ phase shift at this energy using equation~\ref{eq:elastic_qc} yields
\begin{equation*}
	\delta_{\dpi \{\opo\}}(a_t E_{\cm} \approx 0.457) = (9.8 \pm 2.8)^\circ\;.
\end{equation*}
Likewise we can use the second level above threshold in the $L/a_s = 24$ spectrum of the same irrep to constrain the $\dstarpi\{\tpo\}$ phase shift, giving
\begin{equation*}
	\delta_{\dstarpi \{\tpo\}}(a_t E_{\cm} \approx 0.463) = (14.4 \pm 3.4)^\circ\;.
\end{equation*}
These phase shift values characterise the amplitudes near $E_{\dstareta}|_\thr$. Below this energy, the phases follow the typical growth from zero at threshold. While the phase shifts are significantly non-zero they are still small and can be considered in the $K$ matrix using $\gamma$-parameters. 

Furthermore, we have tested in a separate fit including a pole term that the $\dpi$ vector bound state does not affect the $\dpi$ and $\dstarpi$ $1^-$ amplitudes in a quantifiable way above threshold at this light-quark mass where the $D^\ast$ is deeply bound. This is not surprising given that the energy of this state is more than $250$~MeV below the $\dpi$ threshold and $400$~MeV below the $\dstarpi$ scattering region. The same conclusion was reached in ref.~\cite{Moir:2016srx}, where a more detailed treatment of the $1^-$ state can be found.

The $J^P=2^-$ partial wave appears in $[000] E^-$ and $[000] T_2^-$ where only two energy levels can be used. We only require a single channel and can directly determine the phase shift from a single level in $[000] T_2^-$,
\begin{equation*}
	\delta_{\dstarpi \{\tpt\}}(a_t E_{\cm} \approx 0.465) = (1.6 \pm 2.6)^\circ\;,
\end{equation*}
and from a single level in $[000] E^-$,
\begin{equation*}
	\delta_{\dstarpi \{\tpt\}}(a_t E_{\cm} \approx 0.465) = (2.8 \pm 4.5)^\circ\;,
\end{equation*}
which consolidates the previous assumption that the $\dstarpi \{\tpt\}$ contribution is negligibly small in this energy region.

\FloatBarrier

\subsection{Global fit with $K$-matrix}
\label{sec:reference_amplitude}

\begin{figure}[ht]
	\centering
	\includegraphics[width=\textwidth]{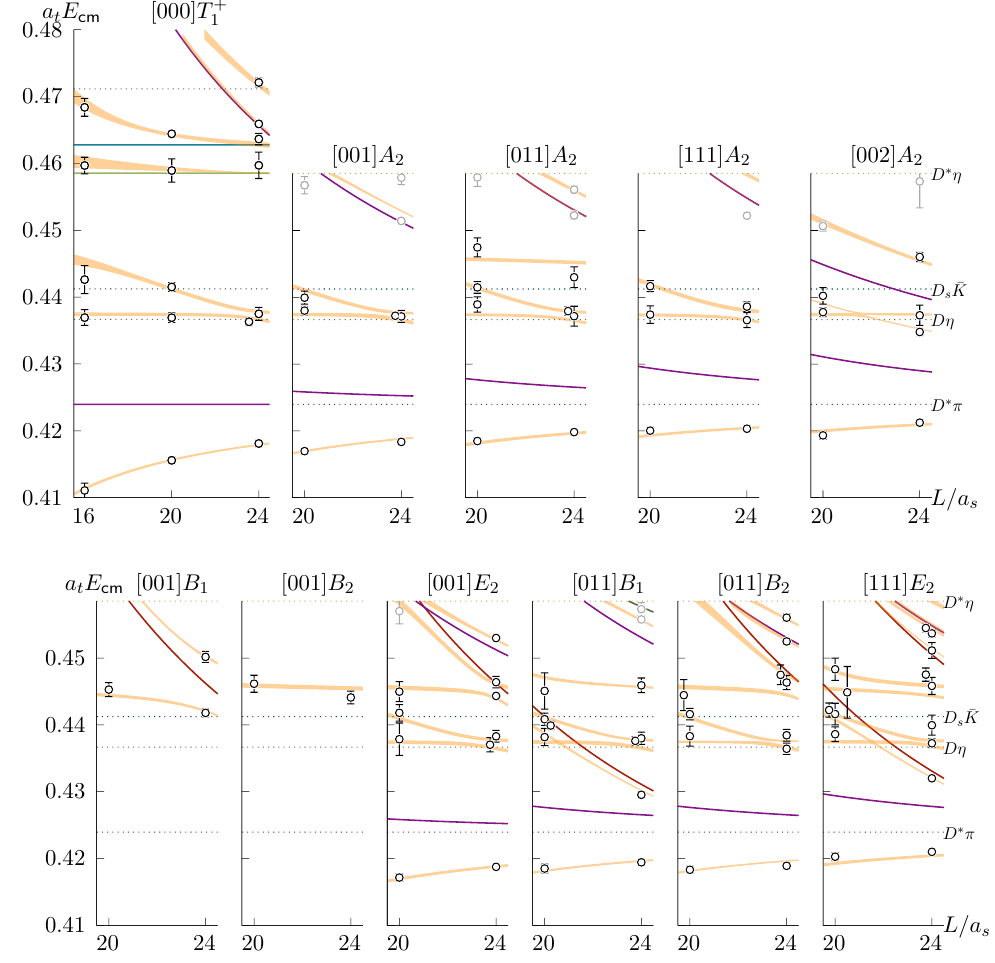}
	\caption[Finite-volume spectra]{Spectra with superimposed lines representing the solution of the finite-volume quantisation condition for the amplitude parametrisation given by the parameters in equation~\ref{eq:kmat_fit_res}.}
	\label{fig:dstarpi:spec_fvs_inelastic}
\end{figure}

A fit to the full dataset is performed using a parametrisation which contains block-diagonal components of angular momentum and parity $1^+$, $1^-$, $2^+$ and $2^-$. When the $0^-$ contribution is allowed to float it is found to be poorly constrained, but relatively small, and we therefore fix it to zero in this reference fit (this choice will be varied later). The $1^+$ part is given by equation~\ref{eq:kmat_coupled}. We allow all channels except $\dstarpi\{\tdo\}$ to couple to the lowest pole. The D-wave coupling to this pole is restricted by its proximity to the threshold, since all partial waves but S-wave are strongly suppressed. The second pole is allowed to couple to $\dstarpi$ $S$ and $D$ wave. Couplings of this pole to $\dstareta$ and $\dsstarkbar$, when included, are again poorly constrained but generally small. We will relax this restriction later. For the smooth part of the amplitude we include parameters for $\dstarpi$ and $\dsstarkbar$ S-wave that are constant in $s$ whereas $\dstarpi$ D-wave and $\dstareta$ S-wave are described by the pole term alone. All of these choices will be varied in section~\ref{subsec:variations}. For the $2^+$ part we use equation~\ref{eq:kmat_2p}. The $1^-$ amplitude is parametrised by a $K$ matrix of constants, given by
\begin{equation}
		{\bm K}_{J^P = 1^-} = 
		\begin{bmatrix}
			\gamma_{\dpi \{\opo\}} & \gamma_{\dpi \{\opo\} \leftrightarrow \dstarpi \{\tpo\}}  \\
			& \gamma_{\dstarpi \{\tpo\}}
		\end{bmatrix}\;.
	\label{eq:kmat_1m}
\end{equation}
Likewise for the $2^-$ a single constant parameter ${\bm K}_{J^P = 2^-} = \gamma_{\dstarpi \{\tpt\}}$ has been included. The parameter values after the $\chi^2$-minimisation and their correlations read
\setcounter{MaxMatrixCols}{20}
\begin{equation*}
	\begin{aligned}[t]
		\begin{array}{rcl}
			g_{0,{D_s^*} \bar{K} \{\tso\}} & = & (-0.337 \pm 0.065  \; { }^{+0.007}_{-0.007}) \cdot a_t^{-1}\\
			g_{0,\dstareta \{\tso\}} & = & (-0.241 \pm 0.097  \; { }^{+0.030}_{-0.035}) \cdot a_t^{-1}\\
			g_{1,\dstarpi \{\tdo\}} & = & (1.72 \pm 0.93  \; { }^{+0.9}_{-1.1}) \cdot a_t\\
			g_{0,\dstarpi \{\tso\}} & = & (0.532 \pm 0.050  \; { }^{+0.007}_{-0.007}) \cdot a_t^{-1}\\
			g_{1,\dstarpi \{\tso\}} & = & (-0.0252 \pm 0.0084  \; { }^{+0.0010}_{-0.0011}) \cdot a_t^{-1}\\
			\gamma_{{D_s^*} \bar{K}\{\tso\}} & = & (-0.72 \pm 0.51  \; { }^{+0.23}_{-0.20}) \\
			\gamma_{\dstarpi\{\tso\}} & = & (2.30 \pm 0.50  \; { }^{+0.36}_{-0.38}) \\
			m_0 & = & (0.42274 \pm 0.00018  \; { }^{+0.00006}_{-0.00010}) \cdot a_t^{-1}\\
			m_1 & = & (0.43753 \pm 0.00023  \; { }^{+0.00010}_{-0.00013}) \cdot a_t^{-1} \vspace{0.3cm}\\
			\gamma_{\dpi\{\opo\}} & = & (16.8 \pm 3.1  \; { }^{+3.6}_{-3.4}) \cdot a_t^2 \\
			\gamma_{\dpi\{\opo\} \leftrightarrow \dstarpi\{\tpo\}} & = & (26 \pm 10  \; { }^{+2}_{-3}) \cdot a_t^2 \\
			\gamma_{\dstarpi\{\tpo\}} & = & (40.2 \pm 6.9  \; { }^{+6.9}_{-5.5}) \cdot a_t^2  \vspace{0.3cm}\\
			g_{0,\dpi \{\odt\}} & = & (1.740 \pm 0.062  \; { }^{+0.017}_{-0.017}) \cdot a_t\\
			g_{0,\dstarpi \{\tdt\}} & = & (1.93 \pm 0.53  \; { }^{+0.38}_{-0.59}) \cdot a_t\\
			\gamma_{\dpi\{\odt\}} & = & (47 \pm 29  \; { }^{+19}_{-18}) \cdot a_t^4 \\
			m_0 & = & (0.44556 \pm 0.00029  \; { }^{+0.00008}_{-0.00007}) \cdot a_t^{-1} \vspace{0.3cm}\\
			\gamma_{\dstarpi\{{}^3 P_2\}} & = & (4.1 \pm 3.8  \; { }^{+5.3}_{-4.9}) \cdot a_t^2 \\
		\end{array}
	\end{aligned}
	\qquad
	\begin{aligned}[t]
		\begin{matrix}
			
			\left.
			\begin{matrix}
				\; \\ \; \\ \; \\ \; \\ \; \\ \; \\ \; \\ \; \\ \;
			\end{matrix}
			\right\} J^P = 1^+ \vspace{0.3cm}\\
			\left.
			\begin{matrix}
				\; \\ \; \\ \;
			\end{matrix}
			\right\} J^P = 1^- \vspace{0.3cm}\\
			\left.
			\begin{matrix}
				\; \\ \; \\ \; \\ \;
			\end{matrix}
			\right\} J^P = 2^+ \vspace{0.3cm}\\
			\left.
			\begin{matrix}
				\;
			\end{matrix}
			\right\} J^P = 2^-\\
		\end{matrix}
	\end{aligned}
\end{equation*}
\resizebox{\textwidth}{!}{
	$\begin{bmatrix}
		1.00 & 0.23 & -0.02 & -0.72 & 0.11 & -0.46 & -0.06 & -0.03 & 0.03 & -0.04 & -0.02 & 0.00 & -0.01 & 0.03 & -0.02 & 0.00 & -0.08\\
		& 1.00 & -0.10 & -0.75 & 0.12 & -0.14 & 0.22 & -0.13 & -0.03 & 0.02 & -0.03 & 0.03 & 0.02 & -0.03 & 0.03 & -0.04 & -0.02\\
		&  & 1.00 & 0.02 & 0.01 & -0.06 & -0.15 & 0.15 & 0.10 & -0.26 & -0.07 & -0.20 & -0.03 & 0.03 & -0.07 & 0.14 & -0.04\\
		&  &  & 1.00 & -0.18 & 0.30 & 0.18 & 0.03 & -0.04 & 0.07 & 0.01 & 0.01 & 0.04 & -0.02 & 0.02 & 0.03 & 0.10\\
		&  &  &  & 1.00 & -0.08 & -0.15 & -0.04 & 0.08 & -0.03 & -0.05 & -0.01 & 0.01 & -0.05 & -0.01 & 0.06 & -0.09\\
		&  &  &  &  & 1.00 & -0.23 & -0.24 & -0.27 & 0.20 & 0.04 & 0.09 & -0.01 & -0.10 & 0.03 & -0.13 & 0.11\\
		&  &  &  &  &  & 1.00 & -0.08 & -0.15 & 0.15 & -0.03 & 0.08 & 0.12 & -0.02 & 0.10 & -0.05 & 0.14\\
		&  &  &  &  &  &  & 1.00 & 0.45 & -0.27 & 0.04 & -0.15 & -0.02 & 0.19 & -0.10 & 0.24 & -0.20\\
		&  &  &  &  &  &  &  & 1.00 & -0.26 & 0.00 & -0.15 & 0.06 & 0.10 & -0.06 & 0.37 & -0.13\\
		&  &  &  &  &  &  &  &  & 1.00 & 0.19 & 0.50 & 0.03 & -0.29 & 0.03 & -0.12 & 0.10\\
		&  &  &  &  &  &  &  &  &  & 1.00 & 0.39 & 0.01 & -0.21 & 0.01 & -0.05 & 0.02\\
		&  &  &  &  &  &  &  &  &  &  & 1.00 & 0.03 & -0.10 & 0.03 & -0.12 & 0.05\\
		&  &  &  &  &  &  &  &  &  &  &  & 1.00 & 0.04 & 0.62 & 0.18 & 0.26\\
		&  &  &  &  &  &  &  &  &  &  &  &  & 1.00 & -0.01 & 0.05 & 0.06\\
		&  &  &  &  &  &  &  &  &  &  &  &  &  & 1.00 & -0.07 & 0.26\\
		&  &  &  &  &  &  &  &  &  &  &  &  &  &  & 1.00 & -0.16\\
		&  &  &  &  &  &  &  &  &  &  &  &  &  &  &  & 1.00\\   
	\end{bmatrix}$
}
\begin{equation}
	\chi^2/\Ndf = \tfrac{100.82}{107 - 17} =  1.12 
	\label{eq:kmat_fit_res}
\end{equation}
The first error quoted is the parameter uncertainty around the $\chi^2$-minimum. The second asymmetric error is based on the variation of the hadron masses (see table~\ref{tab:dstarpi:had_masses}) and the anisotropy within their uncertainties. The nominal value and uncertainty for the latter is $\xi = 3.444 \pm 0.053$. For this purpose we refit this amplitude using an up and down variation of each hadron mass and the anisotropy and then define an envelope around the parameter values obtained in each fit including their statistical uncertainties. This procedure has been used in several other recent works~\cite{Cheung:2020mql,gayer:2021,Lang:2022elg,Wilson:2023hzu,Wilson:2023anv,Yeo:2024chk,Whyte:2024ihh}.
The $\chi^2/\Ndf$-values obtained for the fits corresponding to these variations are given in table~\ref{tab:syst_fits} in the appendix.
The squared amplitude is shown with error bands in figure~\ref{fig:spaghetti}. The energy bands resulting from the solution of the QC for this parametrisation and with the parameter values given in eq.~\ref{eq:kmat_fit_res}, compared to the spectrum obtained from the lattice, is shown in figure~\ref{fig:dstarpi:spec_fvs_inelastic}.

\subsection{Global fit with $SU(3)_F$-symmetric $S$-wave}

Following ref.~\cite{Asokan:2022usm}, we attempt to use an $S$-wave amplitude with manifest SU(3) flavour symmetry.
Motivated by heavy-quark spin symmetry, we use the $\SLJ{1}{S}{0}$ $D\pi-D\eta-D_s\bar{K}$ amplitudes to parameterise the $\SLJ{3}{S}{1}$ $D^\ast\pi-D^\ast\eta-D_s^\ast\bar{K}$ amplitudes.
This amplitude resembles the $K$-matrices used above, but with only overall factors - which may be smooth in $s$ or contain a pole - multiplying constant coupled-channel matrices. These are constructed from the SU(3) factors of the $\overline{\bm{3}}$, ${\bm{6}}$ and the $\overline{\bm{15}}$ to the $D^\ast\pi$-$D^\ast\eta$-$D_s^\ast\bar{K}\SLJc{3}{S}{1}$ channels. The SU(3) amplitude does not contain a $D$-wave. As a simple remedy, we add in a $\dstarpi\{\SLJ{3}{D}{1}\}$ channel that is decoupled from the $S$ wave. This allows for the amplitude to describe the entire spectrum up to our cut-off.
Adding to the parameterisation in eq.~23 of ref.~\cite{Asokan:2022usm}, we use
\begin{align}
	\bm{K}_{J^P=1^+}
	&=
	\left(
	\begin{array}{c@{}c}
		\left[\begin{array}{ccc}
			. & . & .\\
			. & \bm{C}_{\bar{\bm 3}} & .\\
			. & . & .\\
		\end{array}\right] & \mathbf{0}  \\
		\mathbf{0} & 0 \\
	\end{array}\right)
	\left(\frac{g_{\bar{\bm 3}}^2}{m_{\bar{\bm 3}}^2 -s } +  \gamma_{\bar{\bm 3}} \right)
	+\left(
	\begin{array}{c@{}c}
		\left[\begin{array}{ccc}
			. & . & .\\
			. & \bm{C}_{{\bm 6}} & .\\
			. & . & .\\
		\end{array}\right] & \mathbf{0}  \\
		\mathbf{0} & 0 \\
	\end{array}\right)
	{\gamma_{{\bm 6}}} 
	+\left(
	\begin{array}{c@{}c}
		\left[\begin{array}{ccc}
			. & . & .\\
			. & \bm{C}_{\overline{\bm{ 15}}} & .\\
			. & . & .\\
		\end{array}\right] & \mathbf{0}  \\
		\mathbf{0} & 0 \\
	\end{array}\right)
	{\gamma_{\overline{\bm {15}}}} \nonumber
	\\
	&+\left(
	\begin{array}{c@{}c}
		\left[\begin{array}{ccc}
			\;\; & \;\;&\;\; \\
			& \bm{0} & \\
			&  & \\
		\end{array}\right] & \mathbf{0}  \\
		\mathbf{0} & K_{\dstarpi\{\SLJ{3}{D}{1}\}} \\
	\end{array}\right)
	\label{eq:amp_su3}
\end{align}
where the matrix indexing is $1\to D^\ast \pi \SLJc{3}{S}{1}$, $2\to D^\ast \eta \SLJc{3}{S}{1}$, $3\to D_s^\ast \bar{K}\SLJc{3}{S}{1}$ and $4\to D^\ast \pi \SLJc{3}{D}{1}$ and the values of $\bm{C}_{\bar{\bm 3}}$, $\bm{C}_{\bm 6}$ and $ \bm{C}_{\overline{\bm{ 15}}}$ can be found in ref.~\cite{Asokan:2022usm}. We use
\begin{align}
	K_{\dstarpi\{\SLJ{3}{D}{1}\}} = \frac{g_{\SLJ{3}{D}{1}}^2}{m_{\SLJ{3}{D}{1}}^2-s}+\gamma_{\SLJ{3}{D}{1}}^{(0)}+\gamma_{\SLJ{3}{D}{1}}^{(1)}s
\end{align}
for the decoupled $D$-wave, where $m$, $g$, and $\gamma$ are real free parameters, and the appropriate threshold factors are included when eq.~\ref{eq:amp_su3} is inserted in eq.~\ref{eq_tmat}.
The subscripts on the parameters indicate that these affect only the given flavour irrep and partial wave (for $\overline{\bm{3}}$, ${\bm{6}}$ and $\overline{\bm{15}}$ $\SLJ{s}{\ell}{J} = \SLJ{3}{S}{1}$ is implied).
Although the relative strength of each matrix element respects $SU(3)$ symmetry, the differing masses of $\pi,K,\eta$ and $D^\ast,D_s^\ast$ break that symmetry.

Other similar parametrisations were attempted, in particular with additional freedom in the flavour $\bm{6}$. We use a Chew-Mandelstam phase space subtracted at $m_{\bar{\bm 3}}$. The $S$-wave amplitude form used here corresponds to that used in ``fit 4'' of ref.~\cite{Asokan:2022usm}.

Replacing the $J^P=1^+$ part of the reference parameterisation given in eq.~\ref{eq:kmat_fit_res}, we find for the $1^+$ amplitudes,
\begin{align}
	\small
	\setlength{\tabcolsep}{1.5pt}
	\begin{tabular}{rll}
		${m_{\bar{\bm 3}}} = $ & $(0.42277 \pm 0.00017 \pm 0.00007)\cdot a_t^{-1}$ & $\!$\multirow{9}{*}{ $\begin{bmatrix*}[r]   1.00 &  \text{-}0.25 &  \text{-}0.23 &   0.07 &  \text{-}0.29 &   0.30 &   0.08 &   0.06 &   0.06\\
		&  1.00 &   0.42 &   0.47 &  \text{-}0.04 &  \text{-}0.04 &  \text{-}0.05 &  \text{-}0.02 &  \text{-}0.02\\
		&&  1.00 &  \text{-}0.48 &   0.22 &   0.04 &  \text{-}0.16 &  \text{-}0.11 &  \text{-}0.11\\
		&&&  1.00 &  \text{-}0.35 &  \text{-}0.17 &   0.17 &   0.15 &   0.15\\
		&&&&  1.00 &  \text{-}0.13 &  \text{-}0.08 &  \text{-}0.06 &  \text{-}0.06\\
		&&&&&  1.00 &  \text{-}0.57 &  \text{-}0.47 &  \text{-}0.47\\
		&&&&&&  1.00 &   0.82 &   0.81\\
		&&&&&&&  1.00 &   0.36\\
		&&&&&&&&  1.00\end{bmatrix*}$ } \\ 
		$g_{\bm{\bar{3}}} = $ & $(0.75 \pm 0.05 \pm 0.05)\cdot a_t^{-1}$ & \\
		${\gamma_{\overline{\bm {3}}}} = $ & $(6.14 \pm 2.43 \pm 1.44)$ & \\
		${\gamma_{{\bm 6}}}            = $ & $(0.94 \pm 0.55 \pm 0.19)$ & \\
		${\gamma_{\overline{\bm {15}}}}= $ & $(-0.71 \pm 0.60 \pm 0.06)$ & \\
		${m_{\SLJ{3}{D}{1}}} = $           & $(0.43718 \pm 0.00027 \pm 0.00008)\cdot a_t^{-1}$ & \\
		${g_{\SLJ{3}{D}{1}}} = $           & $(7.00 \pm 1.50 \pm 0.09)\cdot a_t$ & \\
		${\gamma^{(0)}_{\SLJ{3}{D}{1}}} = $ & $(62200 \pm 500 \pm 9900)\cdot a_t^{4}$ & \\
		${\gamma^{(1)}_{\SLJ{3}{D}{1}}} = $ & $(-278000 \pm 2000 \pm 44000)\cdot a_t^{6}$ & \\[1.3ex]
		&\multicolumn{2}{l}{ $\chi^2/ N_\mathrm{dof} = \frac{98.9}{107-18} = 1.11$\,,}
	\end{tabular}
\end{align}
where the remaining parameters for $J^P\ne 1^+$ are very similar to those in eq.~\ref{eq:kmat_fit_res}. This amplitude is included in the variations shown in figure~\ref{fig:spaghetti}.

\subsection{Amplitude variations}
\label{subsec:variations}

Prior to an analysis of the analytic structure of these amplitudes in the complex plane, we construct a set of parametrisations similar to eq.~\ref{eq:kmat_fit_res}, based on the $K$-matrix formalism. This reduces to some degree the bias that is introduced by relying on a particular form of parametrisation. The $K$ matrices used are of the general form
\begin{align}
	K_{ij}= \sum_{r} \left\{ \frac{\left(\sum_{n=0}^{N_g} g^{(n)}_{r,i} s^n \right) \; \left(\sum_{n=0}^{N_g} g^{(n)}_{r,j} s^n\right)}{m_r^2-s} \right\} + \sum_{n=0}^{N_{\gamma}} \gamma^{(n)}_{ij} s^n \;,
	\label{eq:kmat_gen}
\end{align}
which allows the coefficients of the pole terms as well as the smooth part of the $K$ matrix to depend polynomially on $s$. We consider two $K$-matrix poles in the $J^P=1^+$ amplitude, and one pole in the $J^P=2^+$ amplitude. The negative-parity amplitudes feature only small interactions that do not require pole terms. 
As in equation~\ref{eq:kmat_coupled}, the indices $i$ and $j$ label the included channels (partial-wave and hadron-hadron), $m_r$ are pole mass parameters and $g^{(n)}_{r,i}$ are coefficients related to the coupling strength of the corresponding channel $i$ to the pole $r$.

\begin{figure}[htb!]
	\includegraphics[width=\textwidth]{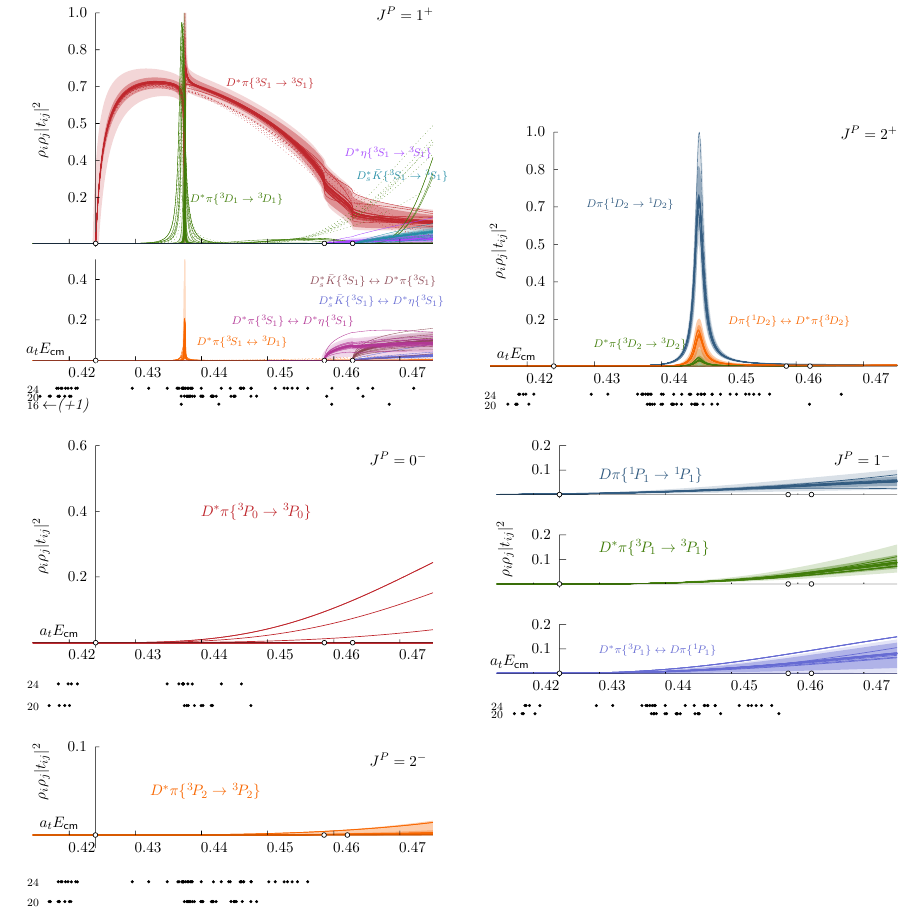}
	\caption{Scattering amplitudes on the real energy axis for $J^+$ (top row) and $J^-$ (middle and bottom row). Partial waves and hadron-hadron channels are distinguished by colour. Each line corresponds to an acceptable amplitude parametrisation. In the case of $J^P = 1^+$ dotted lines indicate parametrisations without the $\dstareta$ and $\dsstarkbar$ channels that have been fitted to a smaller dataset where a cut-off at $E_{\dstareta}|_\thr$ has been applied to the $[000]T_1^+$ spectrum. For the reference parametrisation given by eq.~\ref{eq:kmat_fit_res} two types of error-bands are shown: the darker one expresses the 1$\sigma$ uncertainty around the $\chi^2$ minimum; the lighter one corresponds to the envelope around the line-shapes obtained when varying hadron masses and the anisotropy, including the fit uncertainties, as explained below eq.~\ref{eq:kmat_fit_res}. Below each panel the data constraining the given $J^P$ is shown (without error bars).}
	\label{fig:spaghetti}
\end{figure}

We construct this set of amplitudes by starting from a baseline parametrisation similar to eq.~\ref{eq:kmat_fit_res} and removing or adding parameters. We accepted minimisations with $\chi^2/\Ndf < 1.20$. The full set of amplitudes is given in appendix~\ref{sec:app:amps}. Their line-shapes are shown in figure~\ref{fig:spaghetti}. We group amplitudes by which partial-wave and hadron-hadron channel the variations are applied to (while keeping the others unchanged). In the case of the $J^P = 1^+$ amplitude we also distinguish between variations affecting coupling parameters and the smooth part. This will allow for a more detailed analysis of the effects of parameterisation uncertainties on the analytic continuation in the next section.
We consider polynomials up to linear order for all partial waves except the $\dstarpi \{\tdo\}$ amplitude, where we also allow for a quadratic term. As discussed before, the smooth part of this $D$-wave amplitude is tightly constrained by the spectrum to be zero near $E_{\dsstarkbar}|_\thr$, which in the full-spectrum fit requires either a miniscule coupling parameter $g_{1,\dstarpi \{\tdo\}}$, or a smooth term that is at least linear in $s$. Linear terms often lead to spurious poles in the amplitude which is why we disfavoured them in the reference parameterisation (eq.~\ref{eq:kmat_fit_res}), but we include them in the set of variations to capture the uncertainty that arises for different $K$-matrix choices. In the analysis of the pole content we exclude amplitudes with spurious poles if these are near the poles that we extract. More details on this will be given in section~\ref{sec:poles} and in appendix~\ref{sec:app:amps}. We also include several minimizations that used a lower energy cut-off at $E_{\dstareta}|_\thr$ applied to the $[000]T_1^+$ data. The corresponding parametrisations do not include the heavier channels. Around the narrow $D$-wave resonance these offer a little more variation than the higher energy cut-off.~\footnote{Similarly, in the elastic $D^\ast\pi$ analysis in ref.~\cite{Lang:2022elg}, the tightly-constraining energy level near $E_{\dsstarkbar}|_{\thr.}$ was not included and a larger coupling parameter was obtained, ultimately resulting in a slightly larger pole width. Supplemental section E of that paper also explores this issue, and shows that smaller values can also be obtained with only a small increase in $\chi^2$.} Amplitudes with this lower cut-off are shown with dotted lines in figure~\ref{fig:spaghetti}.

%% file: poles.tex

\section{Poles}
\label{sec:poles}

Partial-wave-projected scattering amplitudes are functions of the Mandelstam variable $s=E_{\cm}^2$. In order to model resonances it is useful to allow $s$ to take complex values, thereby extending the domain of the amplitudes to the entire complex plane, excluding pole singularities and branch cuts. 
Of the branch cuts we only consider those associated with the opening of two-hadron channels. They are a consequence of unitarity and start at the channel threshold extending along the real axis to positive infinity. These cuts are built into the $K$-matrix formalism.
Due to the cuts there is a bifurcation of the complex plane into separate Riemann sheets at every two-hadron threshold such that $n$ open channels lead to $2^n$ sheets. These sheets can be labelled by the sign of the imaginary part of the scattering momenta in the $\cm$-frame, $\left[ \text{sgn} (\Im k_i) \right]_{i = 1, \ldots, n}$ (we order channels by threshold energy from small to large). The sheet given by $\left[\text{sgn} (\Im k_i) = +1\right]$ for $i=1, \ldots, n$ is called the physical sheet since it contains the physical amplitude.
We do not consider left-hand cuts. These arise below the lowest two-hadron threshold due to particle exchanges and extend to negative infinity. 

The motivation for considering unphysical values of $s$ is that it allows bound states and resonances to be identified in a way that does not depend on the parametrisation. These correspond to pole singularities below the lowest scattering threshold or at complex values of $s$ with a non-zero imaginary part.
Close to a $t$ matrix pole, the amplitude is dominated by a term
\begin{equation}
	t_{ij} \sim \frac{c_i c_j}{s_R - s} \;,
	\label{eq:tmat_pole}
\end{equation}
where $i$ and $j$ label the hadron-hadron scattering channels and partial waves that are coupling to the associated state.~\footnote{Kinematic singularities are also possible, which do not correspond to particle states.} The location of the pole can be interpreted as the mass $m_R$ and width $\Gamma_R$ of the resonance or bound state, $s_R = (m_R \pm \tfrac{i}{2} \Gamma_R)^2$. $c_i$ and $c_j$ are residue factors that represent the coupling strength of the channels to the state.
Causality constrains poles on the physical sheet to lie on the real axis below the lowest threshold. These poles are associated with bound states. On all other sheets poles can occur anywhere in the complex plane and are associated with resonances or virtual bound states.
Amplitudes obey the Schwartz-reflection principle, $t_{ij}(s^*) = \left(t_{ij}(s)\right)^*$, such that poles occur in complex-conjugate pairs on the lower and upper half-plane.
If a pole singularity arises due to a physical state coupling to the scattering channel this tends to be a feature that is stable with respect to variations of the amplitude parametrisation. We will therefore focus on poles that appear consistently across our set of parametrisations.

\begin{figure}[htb!]
	\centering
	\begin{subfigure}[l]{1.0\textwidth}
		\includegraphics[width=\textwidth]{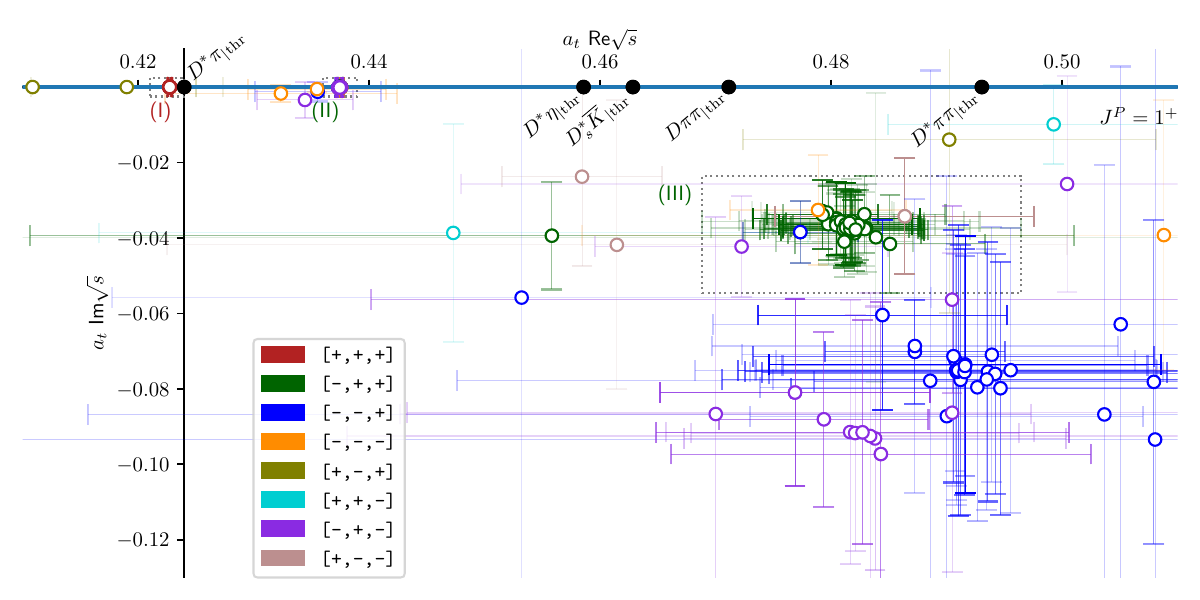}
	\end{subfigure}
	\begin{subfigure}[l]{1.0\textwidth}
		\includegraphics[width=\textwidth]{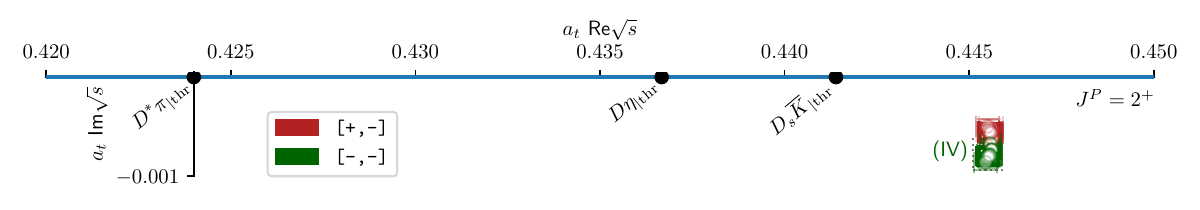}
	\end{subfigure}
	\caption{Top: Poles in $J^P = 1^+$ amplitudes on all Riemann sheets, excluding parametrisations with unphysical singularities. The list of  amplitudes also includes the variations resulting from mass and anisotropy shifts within their respective uncertainties (for the reference parametrisation, eq.~\ref{eq:kmat_fit_res}). Sheets are distinguished by colour and labelled by $\text{sgn}(\Im k_i)$ of the three contributing channels, $\dstarpi$, $\dstareta$ and $\dsstarkbar$, ordered by their masses. The transparency of the error bars representing the fit parameter uncertainties is based on the quality of the fit given by $\chi^2_{\mathrm{min.}}$. Multi-hadron scattering thresholds are represented by filled black circles. The first three pole clusters referenced in the text are marked with dotted rectangles.
		Bottom: Poles in $J^P = 2^+$ amplitudes. Riemann sheets are labelled by $\text{sgn}(\Im k_i)$ of the two channels $\dpi$ and $\dstarpi$, ordered by their masses. The pole cluster on sheet $\left[\texttt{-,-}\right]$ is marked with a dotted rectangle.}
	\label{fig:all_poles}
\end{figure}

Pole singularities in the analytic continuations of the amplitudes are extracted numerically by root-finding the inverse of the $t$ matrix. Parameter uncertainties and correlations are considered by constructing a Monte-Carlo ensemble for each amplitude assuming a Gaussian distribution around the mean values. When clusters of poles are identified in our set of amplitudes we define an envelope around the poles in the cluster including their uncertainties. The same procedure is applied to the corresponding residue factors.
In $J^P=1^+$ there are three open meson-meson channels, $\dstarpi$, $\dstareta$ and $\dsstarkbar$, which split the $\sqrt{s}$~plane into eight Riemann sheets. We consider the energy region of $a_t E_{\dstarpi}|_\thr \lessapprox a_t\Re \sqrt{s} \lessapprox 0.48$ and $0 \geq 2 a_t \Im \sqrt{s} \gtrapprox -a_t m_{\pi}$. Away from thresholds four sheets are relevant in that they smoothly connect to the physical region. These are $\texttt{[+,+,+]}$ below $E_{\dstarpi}|_\thr$ (confined to the real axis), and the lower-half planes of $\texttt{[-,+,+]}$ between $E_{\dstarpi}|_\thr$ and $E_{\dstareta}|_\thr$, $\texttt{[-,-,+]}$ between $E_{\dstareta}|_\thr$ and $E_{\dsstarkbar}|_\thr$, and $\texttt{[-,-,-]}$ above $E_{\dsstarkbar}|_\thr$. Poles on other sheets only affect the amplitude on the real axis significantly if they are close to thresholds, since the path connecting the pole with the physical sheet needs to traverse another sheet. In $J^P = 2^+$ we only consider energies below $E_{\dstareta}|_\thr$ and the channels $\dpi$ and $\dstarpi$. Although $\deta$ and $\dskbar$ are open in this energy region they are strongly suppressed in the tensor channel and therefore neglected. Thus we have four Riemann sheets.
Figure~\ref{fig:all_poles} shows the poles found in all amplitudes including those defined by variations of hadron masses and anisotropy. Clusters of poles marked by dotted rectangles and labelled by Roman numerals will be discussed further below. In $J^P=1^+$ these are located on sheets $\texttt{[+,+,+]}$ and $\texttt{[-,+,+]}$. In $J^P=2^+$ the relevant cluster is found on sheet $\texttt{[-,-]}$.
There are accumulations of poles on other sheets such as $\texttt{[-,-,+]}$, $\texttt{[-,-,-]}$ and $\texttt{[-,+,-]}$ in $J^P = 1^+$ or $\texttt{[+,-]}$ in $J^P = 2^+$. These are either copies of the marked clusters on sheets that are less relevant for the physical amplitude, or the poles do not appear consistently enough or are too far from the real axis to be well-constrained.

\subsection{$J^P = 1^+$ sub-threshold state}

\begin{figure}[htb!]
	\centering
		\includegraphics[width=0.9\textwidth]{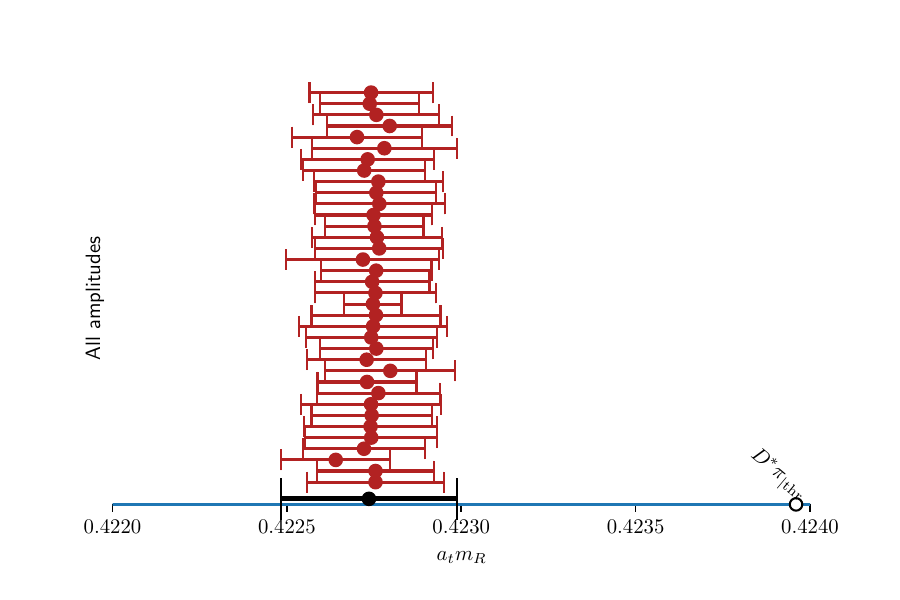}
	\caption{Mass of sub-threshold state ($\mathsf{I}$) found on the real axis of sheet $\texttt{[+,+,+]}$ corresponding to a $J^P=1^+$ bound state for all amplitude parametrisations (red) and envelope around these values corresponding to our final estimate (black). The variations are listed in the tables in appendix~\ref{sec:app:amps}.}
	\label{fig:pole_i_mass}
\end{figure}

\begin{figure}[htb!]
	\centering
		\includegraphics[width=0.95\textwidth]{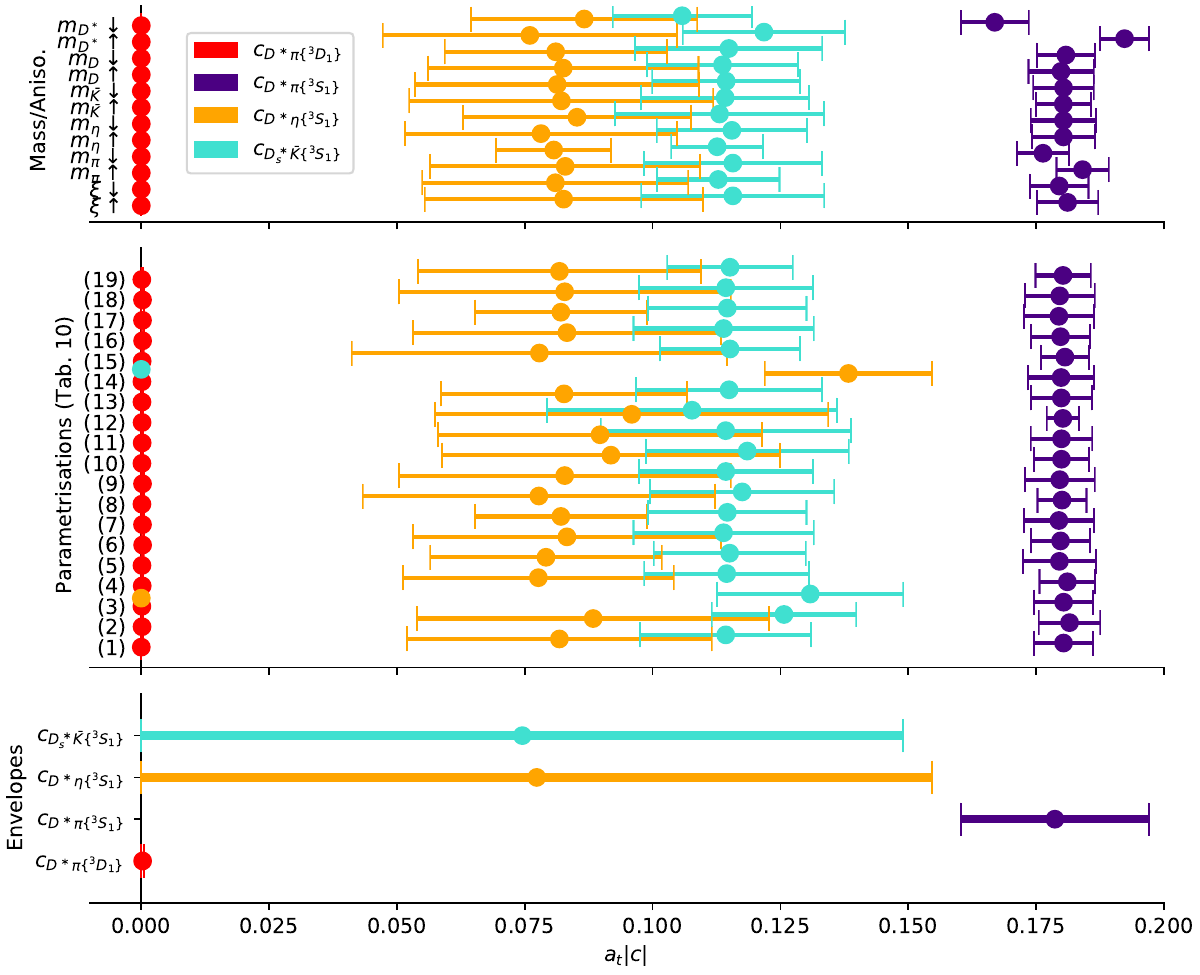}
	\caption{Partial-wave and channel couplings to state ($\mathsf{I}$). We show the couplings for each of the amplitudes corresponding to variations of the stable hadron masses and anisotropy listed in table~\ref{tab:syst_fits} (top), the $1^+$ parametrisation variations listed in table~\ref{tab:amps_1p} (middle), as well as the envelope encompassing all parametrisations and mass/anisotropy variations, including $2^+$ and negative-parity variations given in appendix~\ref{sec:app:amps} but not shown here explicitly, as their effect on these couplings is small (bottom).}
	\label{fig:pole_i_couplings}
\end{figure}

Cluster ($\mathsf{I}$) found on sheet $\left[\texttt{+,+,+}\right]$ corresponds to a shallow bound state. Figure~\ref{fig:pole_i_mass} shows the location of this pole for each amplitude and figure~\ref{fig:pole_i_couplings} the relevant coupling strengths. As can be seen there is good agreement between all parametrisations. The state appears to couple strongly to $\dstarpi$ S-wave and is completely decoupled from D wave. It likely causes the sharp turn-on of the $\tso$ amplitude near the $\dstarpi$ threshold. For the coupling strengths to the heavier meson-meson channels only an upper limit can be determined. This is due to significant freedom in zeroing out coupling coefficients of the lowest-pole term in the K-matrix elements describing the heavier channels, as this pole lies far below the respective thresholds.

\subsection{$J^P = 1^+$ narrow resonance}
\label{subsec:poles_1p_narrow}
\begin{figure}[htb!]
	\centering
	\vspace{-0.5cm}
	\begin{subfigure}[l]{0.9\textwidth}
		\includegraphics[width=\textwidth]{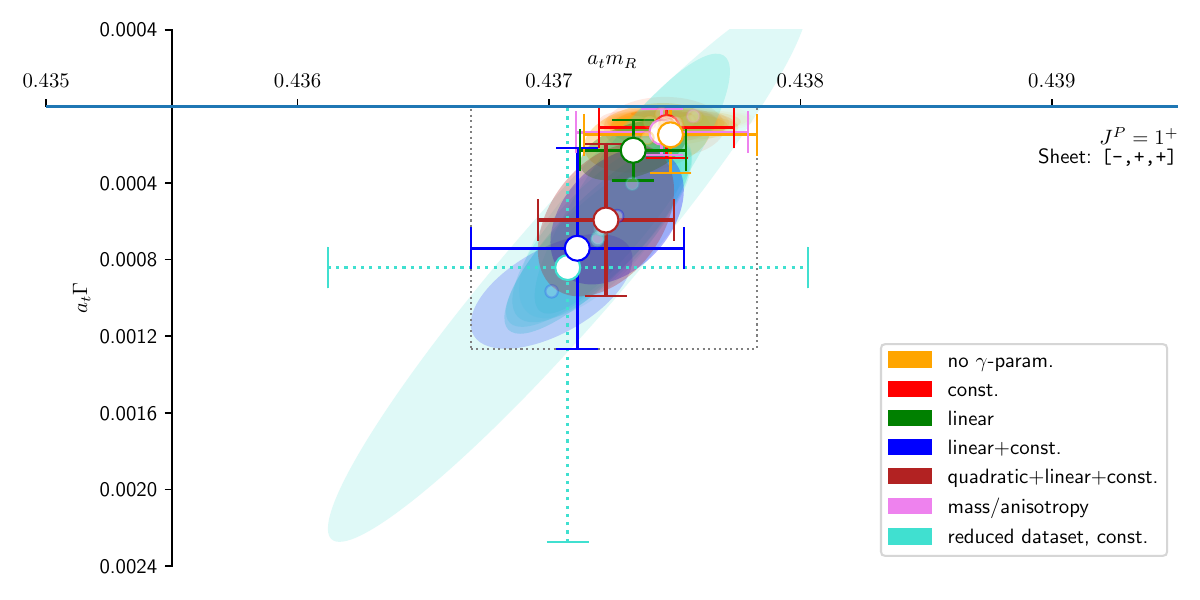}
	\end{subfigure}
	\begin{subfigure}[l]{0.9\textwidth}
		\includegraphics[width=\textwidth]{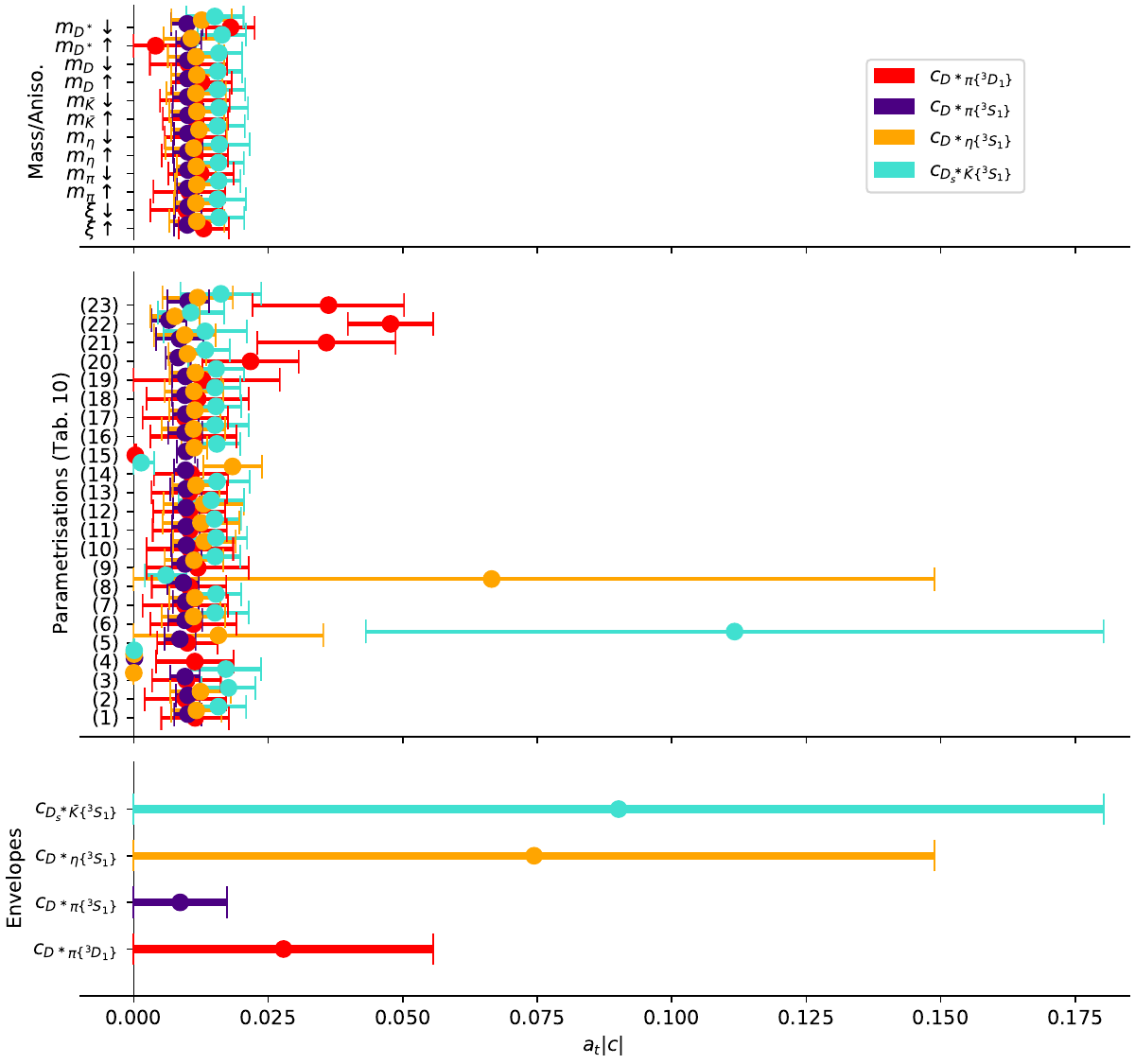}
	\end{subfigure}
	\caption{Top: Pole cluster ($\mathsf{II}$) on sheet $\texttt{[-,+,+]}$ above threshold near the real axis corresponding to a narrow resonance obtained from various sets of parametrisations distinguished by colour. The sets differ in the description of the background part of the D-wave and the dataset used to constrain the fit. Ellipses, representing pole locations and their correlated errors on the mass and width axis, correspond to individual parametrisations, while error bars mark the envelope around the respective set of amplitudes. The transparency of the ellipses is related to the fit quality. The dotted rectangle marks the envelope that defines our final estimate for the mass and width of this state. Note that the reduced-dataset fits are shown for comparison only and are not included in the envelope. Bottom three: as in figure~\ref{fig:pole_i_couplings} but for state ($\mathsf{II}$).}
	\label{fig:pole_ii}
\end{figure}

Cluster ($\mathsf{II}$) is found extremely close to the real axis on sheet $\left[\texttt{-,+,+}\right]$ below the first inelastic threshold. It corresponds to a narrow-resonance state and produces the characteristic peak observed in the $J^P=1^+$ line-shapes in figure~\ref{fig:spaghetti}. Due to its proximity to the real axis and few finite-volume energy levels in the neighbourhood of this resonance it is difficult to precisely determine the couplings and width. For most amplitudes the couplings to all channels are similar in magnitude. The coupling to the closed inelastic channels are very poorly constrained and afflicted with large uncertainties when explicit coupling parameters to these channels are introduced (parametrisations (5) and (8) in figure~\ref{fig:pole_ii} and table~\ref{tab:amps_1p}). The width and couplings depend rather strongly on the parametrisation. This is precisely due to the ambiguity in the $\tdo$ parametrisation that was discussed in section~\ref{sec:scattering}. Figure~\ref{fig:pole_ii} shows that the variation that most strongly affects the width and $\dstarpi\{\tdo\}$ coupling is the polynomial degree of the smooth part of the $\tdo$ amplitude. Linear and higher-order polynomials allow the coupling parameter to take larger values which leads to a larger width. In the same figure, we also include poles of the amplitudes that were fitted to a smaller dataset (table~\ref{tab:elastic_amps}), where a cut-off at the first inelastic threshold was applied to the $[000] T_1^+$ energies. These are shown with dotted error bars and are not included in the envelope that corresponds to our final result. They are compatible with the amplitudes based on the full dataset that posses a linear- or quadratic-order polynomial term. Furthermore our findings suggest that a linear-plus-constant term in the smooth part of the $\tdo$ amplitude provides sufficient flexibility in the fit for the coupling to be non-zero. Indeed figure~\ref{fig:pole_ii} shows that the pole of the amplitude with a quadratic $\tdo$ term does not wander any deeper into the complex plane compared with the linear one.
It should be noted that linear and quadratic-order polynomial terms lead to poles on the physical sheet. Depending on the parametrisation, one is found just below $\text{Re } s = E_{\dstarpi}|_\thr$ around $a_t \text{Im } s \approx -0.035$ and another around $a_t s \approx 0.475 - 0.01 i$ above $E_{\dsstarkbar}|_\thr$. Their couplings to partial waves other than $\dstarpi\{\tdo\}$ are negligibly small. As the physical-sheet poles are far from the narrow resonance it seems permissible to consider these amplitudes to obtain a more careful estimate for the width of this state. We will, however, not consider them in other parts of the analysis.

\subsection{$J^P = 1^+$ broad resonance}

\begin{figure}[htb!]
	\centering
	\begin{subfigure}[l]{1.0\textwidth}
		\includegraphics[width=\textwidth]{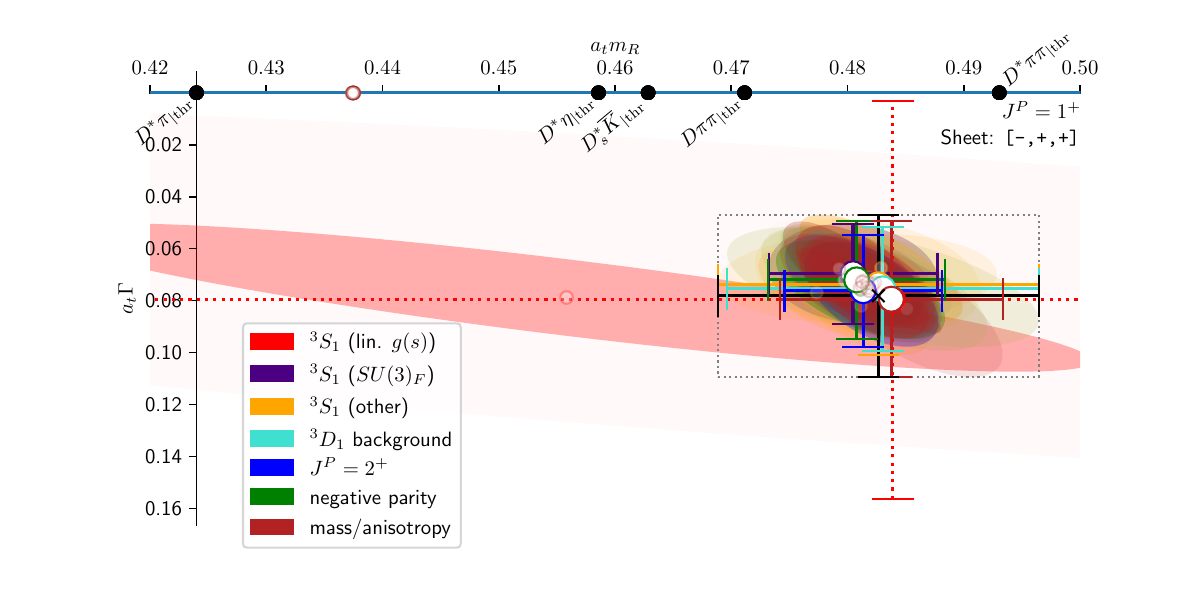}
	\end{subfigure}
	\begin{subfigure}[l]{\textwidth}
		\includegraphics[width=\textwidth]{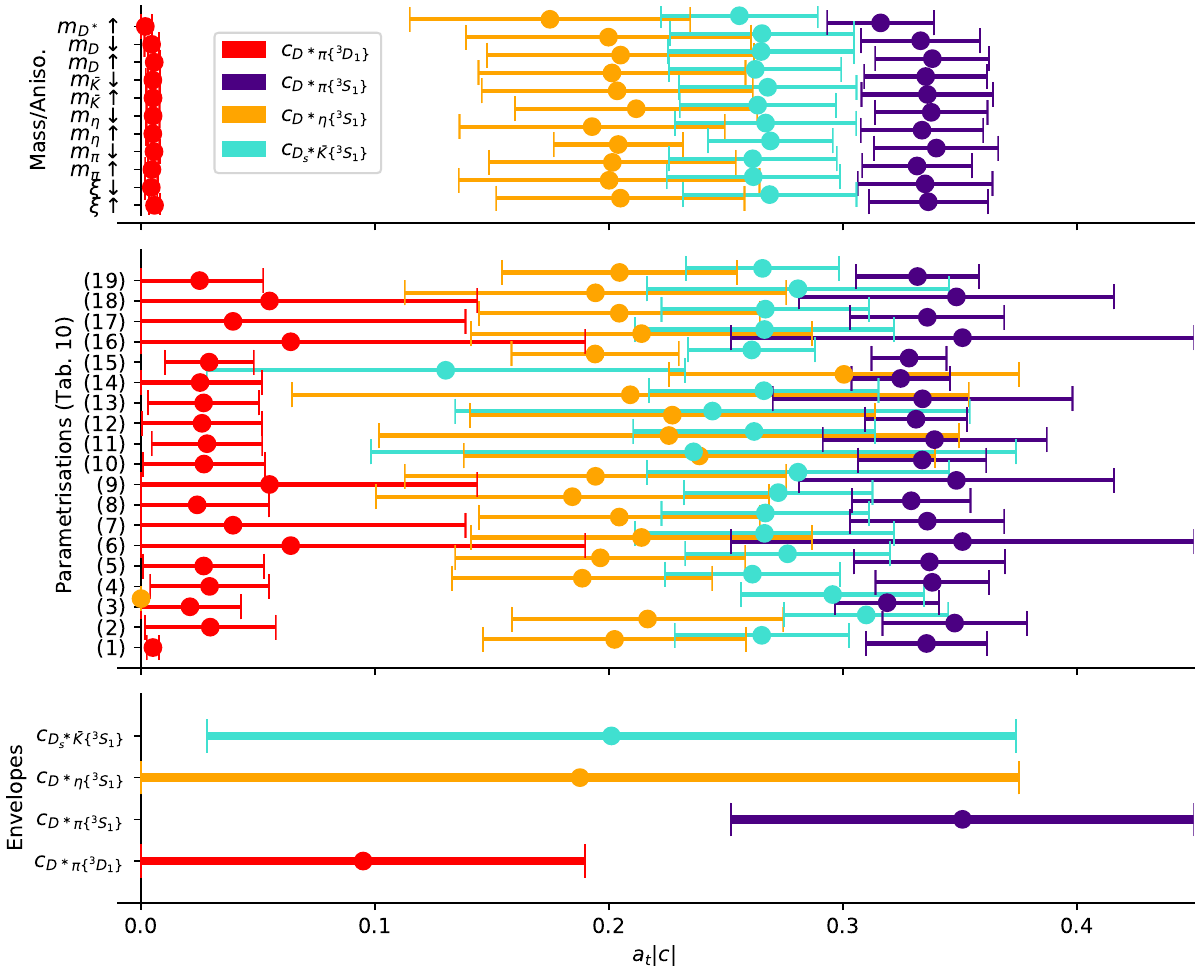}
	\end{subfigure}
	\caption{Top: Pole cluster ($\mathsf{III}$) on sheet $\texttt{[-,+,+]}$ above $E_{\dsstarkbar}|_{\thr}$, corresponding to a broad resonance. Sets of amplitude variations are distinguished by colour. The legend indicates which part of the amplitude has been varied in a given set of parametrisations. Bottom three: as in figure~\ref{fig:pole_i_couplings} but for state ($\mathsf{III}$).}
	\label{fig:pole_iii}
\end{figure}

Cluster ($\mathsf{III}$) is found on sheet $\left[\texttt{-,+,+}\right]$ above the $\dsstarkbar$ threshold and deep in the complex plane (see figure~\ref{fig:pole_iii}). This sheet does not smoothly connect to the physical one at this energy and so this pole does not produce a characteristic resonance peak in the line-shapes of the amplitudes. In fact it can only significantly influence the amplitude around $\dstareta$ threshold. In our parametrisations, this pole is not associated with an explicit $K$ matrix pole-term. 
Its location in the complex plane has large uncertainties. In particular, when considering amplitudes where the coupling term is allowed to depend linearly on $s$, the location of this pole, while still present is poorly constrained. We exclude this type of parametrisation (the corresponding poles are shown with dotted error bars in figure~\ref{fig:pole_iii}) from our estimate.
Nonetheless the existence of the pole can be established in all of the amplitudes. In particular, this pole arises also in the $SU(3)_F$-symmetric amplitude given by eq.~\ref{eq:amp_su3} and is consistent with the other $K$-matrix parametrisations. This is deeper in the complex plane than observed in ref.~\cite{Asokan:2022usm} for the $J^P=0^+$ case, however the pole coupling we find is rather large and so relative importance at real energies may not be so different. 

Excluding linear-order coupling terms, the largest variations of the mass of state ($\mathsf{III}$) are produced by zeroth-order variations of the $\tso$ amplitude including the couplings to the heavier channels and the smooth part, whereas the width is most strongly affected by variations of the hadron masses and anisotropy (specifically the $m_{D^*}$ variation).
The state couples strongly to $\dstarpi\{\tso\}$, and in some cases also to the $S$ wave of the heavier channels, although with large uncertainties.

This pole is located above the upper limit of the energy levels used in this analysis, which also adds to the uncertainty around it. The determination may be improved by going to higher energies in the future, for example by making use of a three-hadron finite-volume quantisation condition~\cite{Briceno:2018aml,Hansen:2019nir,Blanton:2021eyf}.

\FloatBarrier
\subsection{$J^P = 2^+$ narrow resonance}
\begin{figure}[htb!]
	\centering
	\begin{subfigure}[l]{1.0\textwidth}
		\includegraphics[width=\textwidth]{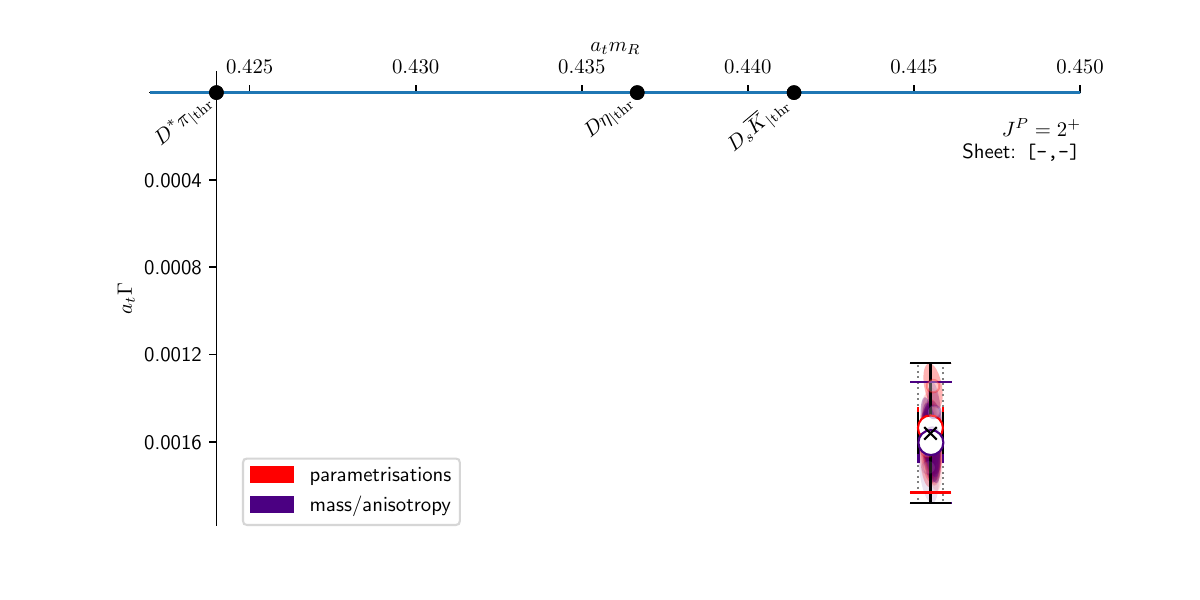}
	\end{subfigure}
	\begin{subfigure}[l]{\textwidth}
		\includegraphics[width=\textwidth]{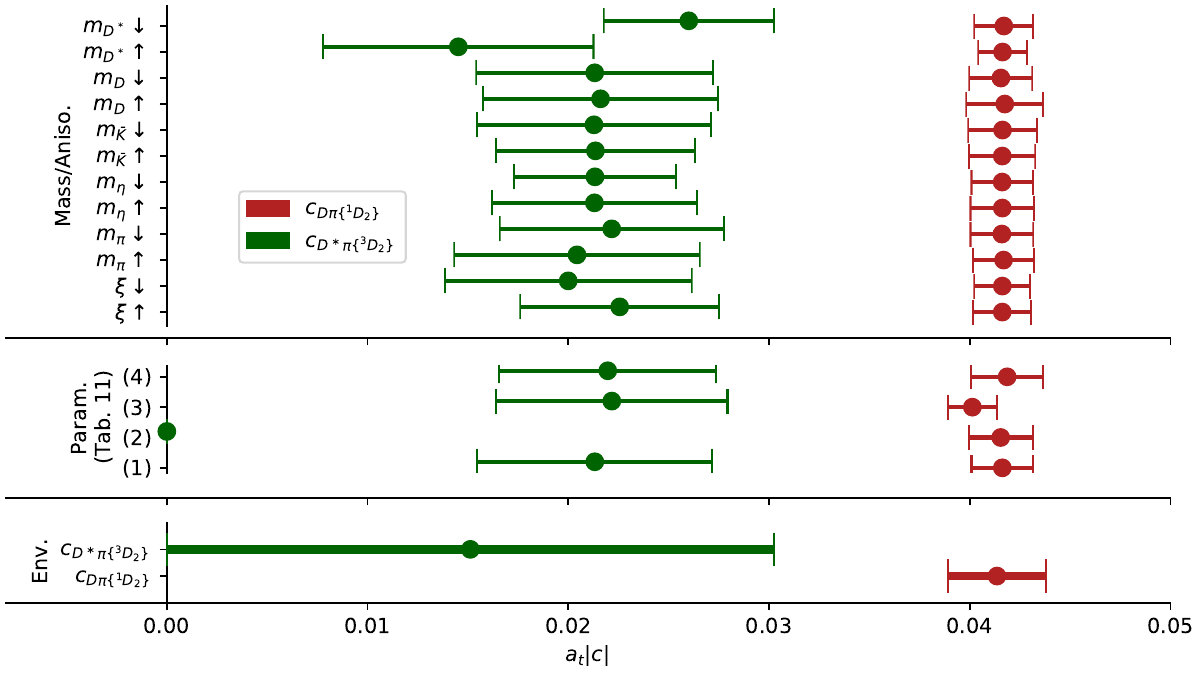}
	\end{subfigure}
	\caption{Top: Pole cluster ($\mathsf{IV}$) on sheet $\texttt{[-,-]}$ in coupled $J^P=2^+$ $\dpi$ and $\dstarpi$ amplitudes, corresponding to a narrow resonance. Sets of amplitude variations are distinguished by colour. Bottom three: as in figure~\ref{fig:pole_i_couplings} but for state ($\mathsf{IV}$). The third panel from the top corresponds to the $2^+$ parametrisation variations listed in table~\ref{tab:amps_2p}.}
	\label{fig:pole_iv}
\end{figure}

In $J^P=2^+$ we find a narrow resonance pole, labelled ($\mathsf{IV}$) in figure~\ref{fig:all_poles} and shown in detail in figure~\ref{fig:pole_iv}, on sheet $\left[\texttt{\texttt{-,-}}\right]$, coupled strongly to $\dpi\{\odt\}$. Most of the parametrisations also exhibit a strong coupling to $\dstarpi\{\tdt\}$, but a vanishing coupling occurs in one parametrisation and cannot be ruled out. The pole being in the vicinity of the physical scattering region on the nearest connected sheet produces the characteristic line-shape of a resonance in the $\dpi\{\odt\}$ (and for most parametrisations the $\dstarpi\{\tdt\}$) amplitude. Copies of this pole exist on sheet $\left[\texttt{\texttt{+,-}}\right]$, slightly closer to the real axis.

\subsection{Summary of results}

In this section we consistently extracted three axial-vector states, one of them a bound state, and two of them resonances. We discuss the possible origin of these in the next section. We also extracted a narrow tensor resonance. The extracted states and their couplings are summarised in tables~\ref{tab:poles} and  \ref{tab:couplings}. We do not rule out additional states as many of our amplitudes exhibit further pole singularities. However, the ones we claim appear to be the most relevant and stable features compatible with our lattice QCD finite-volume energy levels. States ($\mathsf{I}$) and ($\mathsf{IV}$) are fully compatible with our previous result in ref.~\cite{Lang:2022elg}. Regarding state ($\mathsf{II}$) we saw it necessary to replace our earlier estimate for the width with an upper limit seeing that amplitudes with a near-vanishing coupling to this state provide an acceptable description of the data. We had pointed out in the supplemental of ref.~\cite{Lang:2022elg} that such fits converged but were disfavoured by the $\chi^2/\Ndf$ criterion. Given that the added energies that make the most difference are several widths above this resonance, it seems unlikely they should cause a large change to its properties. We consider this new result a slightly more conservative estimate.
State ($\mathsf{III}$) was not found in the earlier publication and is a new result of this analysis.

\begin{table}[htb!]
	\centering
	\begin{tabular}{c|c|c|c|c}
		$J^P$ & sheet & $a_t m_{R}$ & $a_t \Gamma$  & state label \\
		\hline
		\hline
		\multirow{3}{*}{$1^+$} & $\texttt{[+,+,+]}$ & $0.42273(25)$ & -- & ($\mathsf{I}$)  \\
		\cline{2-5}
		& \multirow{2}{*}{$\texttt{[-,+,+]}$} & $0.43726(57)$ & $<0.0013$ & ($\mathsf{II}$) \\
		&  & $0.483(14)$ & $ 0.078(31)$ & ($\mathsf{III}$)  \\
		\cline{2-5}
		\hline
		$2^+$ & $\texttt{[-,-]}$ & $0.44550(39)$ & $0.00156(32)$ & ($\mathsf{IV}$)  \\
	\end{tabular}
	\caption{$J^P = 1^+$ and $2^+$ states found in analytically-continued amplitudes. Masses and widths are determined by an envelope around the poles of all accepted parametrisations, including their uncertainties. The set of amplitudes includes variations of the hadron masses and anisotropies.}
	\label{tab:poles}
\end{table}
\begin{table}[htb!]
	\centering
	\begin{tabular}{c|c|c|c|c}
		$a_t |c_{\dstarpi\{\tso\}}|$  & $a_t |c_{\dstarpi\{\tdo\}}|$  & $	a_t |c_{\dstareta\{\tso\}}|$ & $a_t |c_{\dsstarkbar\{\tso\}}|$ & state label \\
		\hline
		\hline
		$0.179(18)$  & $<0.00051$ & $<0.15$ & $<0.15$ & ($\mathsf{I}$) \\
		$0.0087(86)$ & $<0.056$ & $<0.15$  & $<0.18$ & ($\mathsf{II}$)  \\
		$0.35(10)$ & $<0.19$ & $<0.38$ & $0.20(17)$ & ($\mathsf{III}$)\\
	\end{tabular}
	\vspace{0.35cm}\\
	\begin{tabular}{c|c|c}
		$a_t |c_{\dpi\{\odt\}}|$ & $a_t |c_{\dstarpi\{\tdt\}}|$ & state label \\
		\hline
		\hline
		$0.0414(25)$  &  $<0.030$  & ($\mathsf{IV}$) \\
	\end{tabular}
	\caption{Coupling strengths of the identified $J^P=1^+$ (top) and $2^+$ (bottom) states to scattering channels extracted from factorised residues. The values correspond to the envelope around the individual values of the accepted parametrisations, including their uncertainties.}
	\label{tab:couplings}
\end{table}

%% file: interpretation.tex

\section{Interpretation}
\label{sec:interpretation}

In the previous section we have extracted the locations and residues of four pole singularities in the amplitudes of coupled-channel $J^P=1^+$ $\dstarpi$--$\dstareta$--$\dsstarkbar$ and $J^P=2^+$ $\dpi$--$\dstarpi$ scattering. These describe the masses, widths and couplings of bound and resonant $J^P = 1^+$ states as well as a resonant $J^P = 2^+$ state. In this section we discuss the physical implications of these results and put them in the context of experimental evidence for excited $D$ meson states as well as earlier lattice results and effective field theory calculations.
For context we repeat our main results in scale-set units in table~\ref{tab:results_physical}.
The masses in table~\ref{tab:results_physical} depend on the light-quark mass used and thus do not directly correspond to masses of experimental states. The uncertainties do not include all possible systematic sources, since this calculation is performed at just one lattice spacing.
\begin{table}[htb!]
	\centering
	\bgroup
	\def\arraystretch{1.5}
	\begin{tabular}{c|c|c}
		$J^P$ & state & $ m$ (MeV) \\
		\hline
		\hline
		\multirow{3}{*}{$1^+$} & ($\mathsf{I}$) & $2395.6 \pm 1.4$  \\
		& ($\mathsf{II}$) & $2478.0 \pm 3.2$ \\
		 & ($\mathsf{III}$) & $2737 \pm 79$  \\
		\hline
		$2^+$ & ($\mathsf{IV}$) & $2524.6 \pm 2.2$  \\
	\end{tabular}
	\egroup
	\caption{Masses of lattice states in scale-set units.}
	\label{tab:results_physical}
\end{table}

\subsection{Comparison with $D_1(2430)$, $D_1(2420)$ and $D_2^*(2460)$ results from experiment}

The PDG~\cite{ParticleDataGroup:2024} quotes the masses for the lowest spin-1 and spin-2 D-meson resonances as given in table~\ref{tab:dstarpi:exp_masses}.
A direct comparison with experiment of the masses determined in the lattice computation is not meaningful due to the heavier-than-physical pion mass. Ideally, computations with a range of light quark masses could be performed to understand how this influences the results. However, it is noteworthy that the mass reported by the PDG for the broad $D_1(2430)$, which we may associate with state ($\mathsf{I}$), is larger than our lattice result despite our larger pion mass. This may point towards a similar issue with the interpretation of the experimental results as was found in the case of the scalar $\dzerostar(2300)$. It has been shown (c.f. ref.~\cite{gayer:2021}) that an analysis of the $\dpi\{\osz\}$ amplitude based on a Breit-Wigner fit will yield a BW mass parameter that is significantly larger than the pole mass. We assume that the same conclusion applies here given the similarity of the amplitudes (see section~\ref{subsec:hq_limit}). While state ($\mathsf{I}$) is bound on our lattices we expect it to transition into a broad resonance as the pion mass is lowered. Once the real part of the energy surpasses the two-particle threshold the strong coupling to $\dstarpi$ will generate a large imaginary part as the phase space grows with an increasing gap to the threshold. This expectation is consistent with the experimental observation of a broad amplitude shape.
State ($\mathsf{II}$) may be associated with the narrow $D_1(2420)$. The lattice result is $\approx 55$~MeV above the experimental one, in line with expectations given heavier light quarks. The same is true for state ($\mathsf{IV}$) and the $D_2^*(2460)$, which are $\approx 64$~MeV apart. State ($\mathsf{III}$) does not have a known experimental counterpart. We will discuss its interpretation in section~\ref{sec:inter:two_pole}.
\begin{table}[htb!]
	\centering
	\bgroup
	\def\arraystretch{1.5}
	\begin{tabular}{c|c|c}
		$J^P$ & state & $m$ (MeV) \\
		\hline
		\hline
		\multirow{2}{*}{$1^+$} & $D_1(2430)$  & $(2412 \pm 9)$  \\
		& $D_1(2420)$ & $(2422.1 \pm 0.6)$ \\
		\hline
		$2^+$ & $D_2^*(2460)$ & $2461.1 \pm 0.8$ \\
	\end{tabular}
	\egroup
	\caption[Masses of $D_1$ and $D_2$ mesons]{Masses of the lowest excited $D$ mesons as reported by the PDG~\cite{ParticleDataGroup:2024}}
	\label{tab:dstarpi:exp_masses}
\end{table}

\subsection{Comparison to $D_0^\ast(2300)$ and the heavy-quark limit}
\label{subsec:hq_limit}
\begin{figure}[htb!]
	\centering
	\includegraphics[width=0.7\textwidth]{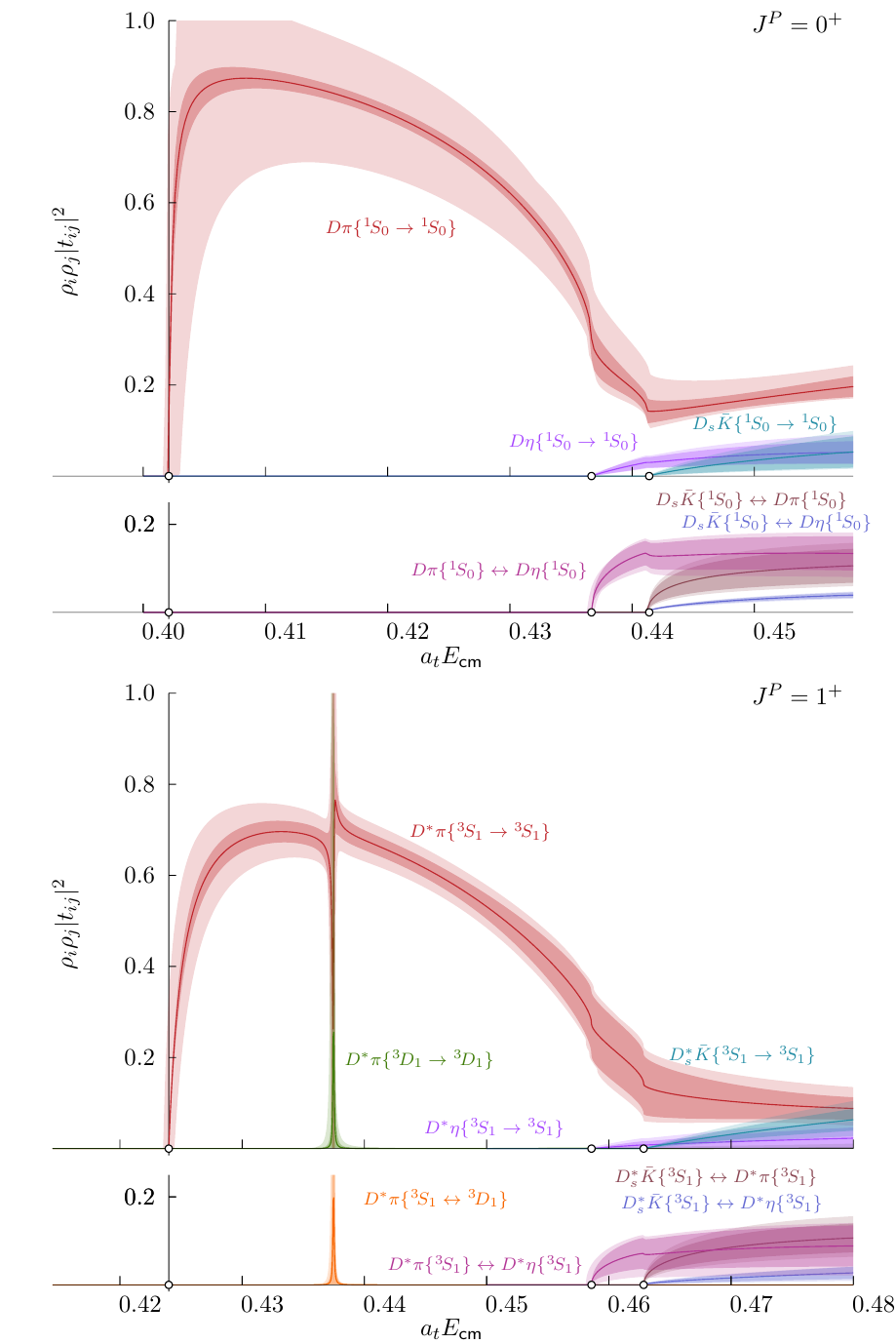}
	\caption{Line-shapes of $J^P=0^+$ $\dpi$, $\deta$ and $\dskbar$ amplitudes from ref.~\cite{Moir:2016srx} (top) and $J^P=1^+$ $\dstarpi$, $\dstareta$ and $\dsstarkbar$ amplitudes from eq.~\ref{eq:kmat_fit_res} (bottom). The outer band represents mass/anisotropy variations, the inner band the statistical uncertainty on the fit parameters.}
	\label{fig:comp_0p1p}
\end{figure}

In the heavy-quark (HQ) limit, $m_Q \rightarrow \infty$, where $Q$ could be either charm or bottom, the dynamics between quarks and gluons are independent of the HQ spin orientation~\cite{DeRujula:1976ugc,Rosner:1985dx,Godfrey:1986wj,Isgur:1991wq,Lu:1991px,Bardeen:2003kt,Godfrey:2005ww,Colangelo:2012xi}.
Thus QCD with infinitely-heavy charm quarks is symmetric under rotations of the charm-quark spin. Furthermore it is always permissible to decompose the total angular momentum in terms of the angular momenta of the light and heavy degrees of freedom, i.e.
\begin{equation}
	J = j_{L} + j_{H} \;.
\end{equation}
If we take $j_H = s_H$ to be the spin of the heavy quark and $j_L$ the sum of the light spins and orbital angular momenta, then a two-fold degeneracy arises for any value of $j_L$ from the two possible orientations $s_{H,z} = \pm \frac{1}{2}$.
Taking $j_L^P = \frac{1}{2}^+$ leads to the degenerate doublet $(J^P) = (0^+, 1^+)$ and $j_L^P = \frac{3}{2}^+$ to $(J^P) = (1^+, 2^+)$. 
In a quark-model picture for instance this would correspond to a light-quark in a $P$-wave orbital around a charm-quark.
The first doublet can be identified with the $D_0^\ast(2300)$ and $D_1(2430)$, the second with the $D_1(2420)$ and $D_2^*(2460)$.
The mass differences within the doublets should be due to corrections of $\mathcal{O}(1/m_c)$.
The HQ limit suggests that the signatures in the $S$ and $D$ wave scattering amplitudes of states belonging to a doublet should be very similar, which matches what we observe in this calculation. In figure~\ref{fig:comp_0p1p} we compare the $S$ wave amplitudes of $\dpi$ ($\deta$, $\dskbar$) (based on the result in ref.~\cite{Moir:2016srx}) and $\dstarpi$ ($\dstareta$, $\dsstarkbar$) scattering (eq.~\ref{eq:kmat_fit_res}) constrained by energies from the same lattice ensembles corresponding to $m_\pi = 391$~MeV. In the scalar case the pole, corresponding to the $\dzerostar(2300)$, is statistically indistinguishable from the $\dpi$ threshold, leading to a sharper turn-on of the amplitude. Barring the region of the narrow $1^+$ resonance, which introduces a disturbance in the $\dstarpi\{\tso\}$ amplitude due to the dynamical coupling of $S$ and $D$ wave, the amplitudes show otherwise very similar behaviour, including at the $D^{(*)}\eta$ and $D_s^{(*)}\bar{K}$ thresholds. The same is true for the narrow resonances in the $\dstarpi\{\tdo\}$ and $\dstarpi\{\tdt\}$ amplitudes.
A prediction of HQ spin-symmetry is that one of the $1^+$ states decays exclusively to $\tso$. The couplings extracted for state ($\mathsf{I}$) support this prediction. The other $1^+$ state is expected to decay exclusively to $\tdo$. The lattice result for state ($\mathsf{II}$) is not conclusive regarding this expectation but the limits on our couplings are not incompatible with HQ spin-symmetry.

\subsection{The two-pole structure of the $S$-wave amplitude}
\label{sec:inter:two_pole}

The existence of a second pole above the ground-state scalar in $\dpi$ ($\deta$, $\dskbar$) and ground-state axial-vector in $\dstarpi$ ($\dstareta$, $\dsstarkbar$) $S$-wave amplitudes was first predicted using unitarised chiral perturbation theory (u$\chi_{\text{PT}}$). When considering the SU(3)-flavour decomposition of the scattering amplitude of the pseudoscalar meson octet with the triplet of charmed mesons, $\bm{\overline{3}}\otimes {{\bm 8}} \to {\overline{\bm 3}}\oplus {{\bm 6}}\oplus \overline{\bm{15}}$, a pole arises due to interactions in the flavour-sextet irrep.
Most of the relevant calculations have been performed in the scalar sector~\cite{Kolomeitsev:2003ac,Hofmann:2003je,Gamermann:2006nm,Guo:2006fu,Guo:2009ct,Doring:2011vk,ALBALADEJO2017465,Du:2017zvv,Meissner:2020khl}. 
But given HQ symmetry, any results of higher poles in the scalar sector are also relevant for the axial-vector. 
There are a few studies that explicitly consider the $1^+$ channel~\cite{gamermannAxialResonancesOpen2007,guoDynamicallyGeneratedHeavy2007,Du:2017zvv,guoChiralExcitationsExotic2019b,Meissner:2020khl}.
References~\cite{Du:2017zvv,Meissner:2020khl} discuss both the scalar and axial-vector heavy-light resonances. Reference~\cite{Du:2017zvv} gives $(2555\substack{+47 \\ -30} - 203\substack{+8 \\ -9} i)$~MeV for the mass and half-width of the higher pole in the $J^P=1^+$ $S$-wave amplitude. The width of this state is compatible with lattice state ($\mathsf{III}$) while the mass is about $200$~MeV below our result, which is not unexpected given the differing light-quark/pion masses.

In ref.~\cite{guoChiralExcitationsExotic2019b}  a $J^P=1^+$ pole is found at $\sqrt{s} = (2606\substack{+23 \\ -30 } - 59\substack{+13 \\ -25 }i)$~MeV (uncertainties based on the variation of the matching scale) on sheet $\texttt{[-,-,+]}$ for their preferred set of fit parameters in a calculation with physical meson masses. There are indications of a pole on this sheet and roughly compatible in mass in our amplitudes, albeit much deeper in the complex plane. We do not make any precise claims about the mass and width of this state because the combined uncertainties on this cluster of poles are too large.
In the context of the $\dzerostar(2300)$, ref.~\cite{ALBALADEJO2017465} presents a u$\chi_{\text{PT}}$ $J^P=0^+$ amplitude that is consistent with the lattice energy levels extracted in ref.~\cite{Moir:2016srx} (where the same ensemble is used as in this study). While imposing chiral symmetry on the $J^P = 0^+$ amplitudes gives rise to a sextet pole at $(2468\substack{+32 \\-25 } - 113\substack{+18 \\ -16 } i)$~MeV (for $m_\pi = 391$~MeV) this pole was not reported in ref.~\cite{Moir:2016srx}, based only on simple $K$-matrix amplitudes. No claim of a higher $0^+$ pole was made, although poles were found around or beyond the highest energy levels used to determine the amplitudes.
 Reference~\cite{Asokan:2022usm} re-analyses the rest-frame lattice energies of ref.~\cite{Moir:2016srx} using flavour SU(3)-symmetric K-matrix amplitudes. They find a pole consistent with the u$\chi_{\text{PT}}$ results. This suggests that the location of this pole may quite strongly depend on the parametrisation and this conclusion is also reached in ref.~\cite{Asokan:2022usm}. This logic would equally apply in the axial-vector case. Indeed we saw indications that the mass of state ($\mathsf{III}$) may be different if coupling parameters are allowed to depend on the energy. However, despite attempting a fit with a flavour-SU(3)-symmetric $K$-matrix (eq. \ref{eq:amp_su3}) we did not find a pole at a lower energy or closer to the real axis as suggested by ref.~\cite{Asokan:2022usm} for the scalar sector, but instead one that is consistent with our other $K$-matrices.

For $J^P=0^+$, a virtual bound state in the flavour-$\bm{6}$ channel has recently been found using lattice QCD in another analysis by the Hadron Spectrum Collaboration at $m_\pi \approx 700$~MeV, where the light and strange quark masses are set equal, giving exact SU(3) flavour symmetry~\cite{Yeo:2024chk}. This state is also found in the $(S,I) = (-1,0)$ component of $D\bar{K}$ scattering in ref.~\cite{Cheung:2020mql}. By HQ symmetry these findings suggest the existence of a similar state in the corresponding axial-vector amplitudes. Hence the existence of this pole is not in doubt. However, the question about its precise location appears to be delicate, in particular for the $I=1/2$ cases where coupled-channel scattering amplitudes are needed.

%% file: summary.tex
\section{Summary}
\label{sec:summary}

We presented the first lattice calculation of coupled $\dstarpi$, $\dstareta$ and $\dsstarkbar$ scattering in the axial-vector channel and coupled $\dpi$ and $\dstarpi$ scattering in the tensor channel. Using a finite-volume quantisation condition, the scattering amplitudes have been fitted to 107 lattice energy levels computed over a large basis of  interpolating operators with quantum numbers $I=\frac{1}{2}, S=0, C=1$, on lattices of three different volumes with quark-masses corresponding to $m_\pi = 391$~MeV. The results are broadly consistent with our earlier study of elastic axial-vector $\dstarpi$ scattering~\cite{Lang:2022elg}. The $\dstarpi\{\tso\}$ amplitude turns on sharply at threshold and remains large throughout the elastic scattering region. The S-waves of the heavier channels turn on rapidly as well at their respective thresholds but are significantly smaller. The $\dstarpi\{\tdo\}$ amplitude has the shape of a narrow resonance between the $\dstarpi$ and $\dstareta$ thresholds. It proved difficult to extract a precise width of this resonance. Additional lattice volumes would provide the necessary energy levels to better constrain this amplitude. Mixing between the channels and some amount of mixing between partial waves is found. The tensor-channel $\dpi\{\odt\}$ and $\dstarpi\{\tdt\}$ amplitudes also exhibit the shape of a narrow resonance and there is mixing between these channels.

The analytic continuation of the amplitudes reveals two pole singularities of $J^P=1^+$ below the $\dstareta$ threshold, corresponding, in scale-set units, to a bound-state at $(2395.6\pm1.4)$~MeV and a narrow resonance at $(2478.0\pm3.2)$~MeV, and one pole singularity of $J^P=2^+$ corresponding to a narrow resonance at $(2524.6\pm2.2)$~MeV. These states may be qualitatively compared to the experimentally established $D_1(2430)$, $D_1(2420)$ and $D_2^*(2460)$.
Notably the lowest $J^P=1^+$ state is found lower in mass than the experimental counterpart, the $D_1(2430)$, despite using heavier light-quarks in the lattice QCD calculation. This might be traced to a disparity between the Breit-Wigner mass parameter and pole location in this amplitude, that has already been noted in the scalar sector.
Furthermore, we find an additional $J^P=1^+$ resonance above the $\dsstarkbar$ threshold at $(2737\pm79)$~MeV and with a substantial width, which strongly couples to $\dstarpi\{\tso\}$. This pole may likely be related to the conjectured flavour-sextet pole and supports the claims of a two-pole structure of the S-wave amplitude.
Although the pole was found above our highest energy level, it was present in every amplitude, and so appears to be a robust feature. While the existence of this pole, both in this work and related studies, can be claimed with confidence, its location may depend strongly on the amplitude parametrisation. Indications of other poles on more remote sheets are also seen, but these are likely less important for scattering at real energies.

\begin{figure}[htb!]
	\centering
	\graphicspath{{plots/}}
	\includegraphics[width=\textwidth]{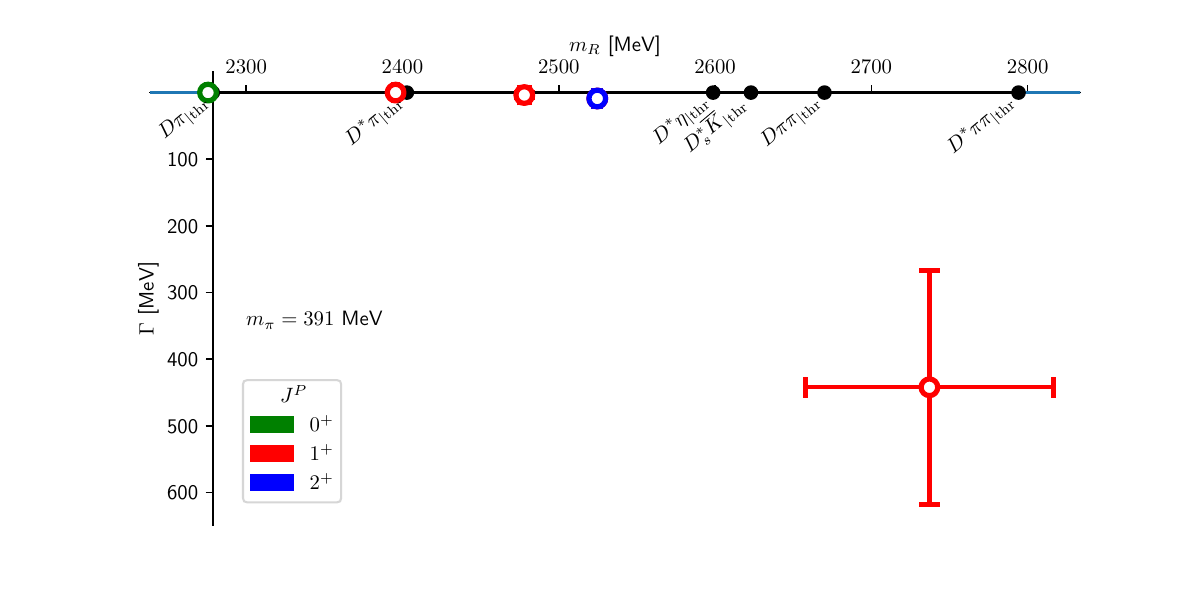}
	\caption{Summary of $J^P=0^+$, $1^+$ and $2^+$ states found in this calculation, where $m_\pi = 391$~MeV. Points with error bars represent our final estimates of the masses and widths of the states, their colour distinguishing the values of $J^P$. Black filled circles on the real axis represent relevant thresholds of two- and three-meson scattering.}
	\label{fig:summary}
\end{figure}

Comparing to $J^P=0^+$ results from previous lattice studies, we find a strong signature of HQ spin-symmetry in the amplitudes. The extracted couplings are roughly consistent with expectations in the HQ limit. We summarise our results in scale-set units in figure~\ref{fig:summary} where we also include the $\dzerostar$ from ref.~\cite{Moir:2016srx}.
To better understand the higher $1^+$ pole a lattice computation like ref.~\cite{Yeo:2024chk} at the flavour-$SU(3)$-symmetric point, but for the axial-vector channel, could be performed. Also a computation at an intermediate mass point, where $SU(3)$ symmetry is very weakly broken, may provide valuable insights. Furthermore, to extend the energy range of our calculation and properly constrain the amplitude beyond the $\dsstarkbar$ threshold, the recently-developed three-body formalism~\cite{Briceno:2024ehy,Draper:2024qeh,Jackura:2023qtp,Blanton:2019vdk,Blanton:2020gmf,Fischer:2020jzp,Hammer:2017kms} will be needed, as $D \pi \pi$ and $D^* \pi \pi$ contributions become relevant. This would also be the case for a computation with lighter light-quarks, that would allow for a robust extrapolation to the physical point. Furthermore, a search for the analogue of these states with $b$ quarks would be an interesting investigation.

%% file: ack.tex
\begin{acknowledgments}

We thank our colleagues within the Hadron Spectrum Collaboration (www.hadspec.org), in particular Christopher Thomas and Daniel Yeo for comments on the manuscript.
NL acknowledges support by the EU H2020 research and innovation programme under the Staff Exchange grant agreement No-101086085-ASYMMETRY, as well as by the Spanish Ministerio de Ciencia e Innovacion project PID2020-113644GB-I00  and by Generalitat Valenciana through the grant PROMETEO/2019/083.
{DJW acknowledges support from a Royal Society University Research Fellowship, and support from the U.K. Science and Technology Facilities Council (STFC) [grant numbers ST/T000694/1 \& ST/X000664/1].}
The software codes {\tt Chroma}~\cite{Edwards:2004sx}, {\tt QUDA}~\cite{Clark:2009wm,Babich:2010mu}, {\tt QPhiX}~\cite{Joo:2013lwm}, and {\tt QOPQDP}~\cite{Osborn:2010mb,Babich:2010qb} were used to compute the propagators required for this project.
This work used the Cambridge Service for Data Driven Discovery (CSD3), part of which is operated by the University of Cambridge Research Computing Service (www.csd3.cam.ac.uk) on behalf of the STFC DiRAC HPC Facility (www.dirac.ac.uk). The DiRAC component of CSD3 was funded by BEIS capital funding via STFC capital grants ST/P002307/1 and ST/R002452/1 and STFC operations grant ST/R00689X/1. Other components were provided by Dell EMC and Intel using Tier-2 funding from the Engineering and Physical Sciences Research Council (capital grant EP/P020259/1).
This work also used clusters at Jefferson Laboratory under the USQCD Initiative and the LQCD ARRA project, and the authors acknowledge support from the U.S. Department of Energy, Office of Science, Office of Advanced Scientific Computing Research and Office of Nuclear Physics, Scientific Discovery through Advanced Computing (SciDAC) program, and the U.S. Department of Energy Exascale Computing Project.
This research was supported in part under an ALCC award, and used resources of the Oak Ridge Leadership Computing Facility at the Oak Ridge National Laboratory, which is supported by the Office of Science of the U.S. Department of Energy under Contract No. DE-AC05-00OR22725. This research is also part of the Blue Waters sustained-petascale computing project, which is supported by the National Science Foundation (awards OCI-0725070 and ACI-1238993) and the state of Illinois. Blue Waters is a joint effort of the University of Illinois at Urbana-Champaign and its National Center for Supercomputing Applications. This work is also part of the PRAC “Lattice QCD on Blue Waters”. This research used resources of the National Energy Research Scientific Computing Center (NERSC), a DOE Office of Science User Facility supported by the Office of Science of the U.S. Department of Energy under Contract No. DEAC02-05CH11231. The authors acknowledge the Texas Advanced Computing Center (TACC) at The University of Texas at Austin for providing HPC resources that have contributed to the research results reported within this paper.
Gauge configurations were generated using resources awarded from the U.S. Department of Energy INCITE program at Oak Ridge National Lab, NERSC, the NSF Teragrid at the Texas Advanced Computer Center and the Pittsburgh Supercomputer Center, as well as at Jefferson Lab.

\end{acknowledgments}

%% file: appendix_operators.tex
\section{Operator Lists}
\label{app:sec:ops}

The meson-meson operators and the number of $\bar{q}q$-like operators used in the computation of the spectra discussed in section~\ref{sec:lattice} are listed in table~\ref{tab:app:dstarpi:ops2}. The operators are grouped by irrep and momentum.

\begin{table}[!htb]
	\centering
	\resizebox{\textwidth}{!}{
		\begin{tabular}{lllllll}
			\toprule
			$[000] T_1^+$ & $[000]E^+$ & $[000]T_2^+$ & $[001]A_2$ & $[001] E_2$ & $[001] B_1$ & $[001] B_2$ \\
			\midrule
			$D_{[000]} \rho_{[000]}$ (1) & $D_{[100]} \pi_{[100]}$ (1) & $D_{[110]} \pi_{[110]}$ (1) & $D_{[100]} \rho_{[000]}$ (1) & $D_{[100]} \pi_{[110]}$ (1) & $D_{[100]} \pi_{[110]}$ (1) & $D_{[111]} \pi_{[110]}$ (1) \\
			$D_{[100]} \rho_{[100]}$ (2) & $D_{[110]} \pi_{[110]}$ (1) & ${D^*}_{[100]} \pi_{[100]}$ (1) & $D_{[000]} {f_0}_{[100]}$ (1) & $D_{[110]} \pi_{[100]}$ (1) & $D_{[110]} \pi_{[100]}$ (1) & $D_{[110]} {f_0}_{[100]}$ (1) \\
			$D_{[100]} {f_0}_{[100]}$ (1) & $D_{[200]} \pi_{[200]}$ (1) & ${D^*}_{[000]} \rho_{[000]}$ (1) & $D_{[100]} {f_0}_{[000]}$ (1) & $D_{[111]} \pi_{[110]}$ (1) & $D_{[100]} \eta_{[110]}$ (1) & ${D^*}_{[100]} \pi_{[110]}$ (2) \\
			${D^*}_{[000]} \pi_{[000]}$ (1) & $D_{[100]} \eta_{[100]}$ (1) & $\bar{q} \mathbf{\Gamma} q$ (29) & ${D^*}_{[000]} \pi_{[100]}$ (1) & $D_{[110]} \eta_{[100]}$ (1) & $D_{[110]} \eta_{[100]}$ (1) & ${D^*}_{[110]} \pi_{[100]}$ (2) \\
			${D^*}_{[100]} \pi_{[100]}$ (2) & $D_{[110]} \eta_{[110]}$ (1) &  & ${D^*}_{[100]} \pi_{[000]}$ (1) & $D_{[000]} \rho_{[100]}$ (1) & ${D_s}_{[100]} \bar{K}_{[110]}$ (1) & $\bar{q} \mathbf{\Gamma} q$ (20) \\
			${D^*}_{[110]} \pi_{[110]}$ (3) & ${D_s}_{[100]} \bar{K}_{[100]}$ (1) &  & ${D^*}_{[110]} \pi_{[100]}$ (2) & $D_{[100]} \rho_{[000]}$ (1) & ${D_s}_{[110]} \bar{K}_{[100]}$ (1) &  \\
			${D^*}_{[000]} \eta_{[000]}$ (1) & ${D_s}_{[110]} \bar{K}_{[110]}$ (1) &  & ${D^*}_{[100]} \eta_{[000]}$ (1) & ${D_s}_{[110]} \bar{K}_{[100]}$ (1) & $\bar{q} \mathbf{\Gamma} q$ (12) &  \\
			${D^*}_{[100]} \eta_{[100]}$ (2) & $\bar{q} \mathbf{\Gamma} q$ (4) &  & ${D_s^*}_{[100]} \bar{K}_{[000]}$ (1) & ${D^*}_{[000]} \pi_{[100]}$ (1) &  &  \\
			${D^*}_{[000]} \rho_{[000]}$ (1) &  &  & ${D_0}_{[100]} \pi_{[000]}$ (1) & ${D^*}_{[100]} \pi_{[000]}$ (1) &  &  \\
			${D_s^*}_{[000]} \bar{K}_{[000]}$ (1) &  &  & $\bar{q} \mathbf{\Gamma} q$ (32) & ${D^*}_{[110]} \pi_{[100]}$ (3) &  &  \\
			${D_s^*}_{[100]} \bar{K}_{[100]}$ (2) &  &  &  & ${D^*}_{[000]} \eta_{[100]}$ (1) &  &  \\
			${D_0}_{[100]} \pi_{[100]}$ (1) &  &  &  & ${D^*}_{[100]} \eta_{[000]}$ (1) &  &  \\
			$\bar{q} \mathbf{\Gamma} q$ (44) &  &  &  & ${D^*}_{[100]} {f_0}_{[000]}$ (1) &  &  \\
			&  &  &  & ${D_s^*}_{[100]} \bar{K}_{[000]}$ (1) &  &  \\
			&  &  &  & $\bar{q} \mathbf{\Gamma} q$ (44) &  &  \\
			\bottomrule
		\end{tabular}
	}
	
	\resizebox{\textwidth}{!}{
		\begin{tabular}{lllllll}
			\toprule
			$[000] T_1^-$ & $[011] A_2$ & $[011] B_1$ & $[011] B_2$ & $[111] A_2$ & $[111] E_2$ & $[002] A_2$ \\
			\midrule
			$D_{[100]} \pi_{[100]}$ (1) & $D_{[110]} \pi_{[110]}$ (1) & $D_{[100]} \pi_{[100]}$ (1) & $D_{[110]} \pi_{[110]}$ (1) & $D_{[111]} \rho_{[000]}$ (1) & $D_{[100]} \pi_{[110]}$ (1) & $D_{[100]} \rho_{[100]}$ (1) \\
			$D_{[100]} \eta_{[100]}$ (1) & $D_{[110]} \rho_{[000]}$ (1) & $D_{[110]} \pi_{[110]}$ (1) & $D_{[111]} \pi_{[100]}$ (1) & $D_{[111]} {f_0}_{[000]}$ (1) & $D_{[110]} \pi_{[100]}$ (1) & $D_{[100]} {f_0}_{[100]}$ (1) \\
			${D^*}_{[100]} \pi_{[100]}$ (1) & $D_{[110]} {f_0}_{[000]}$ (1) & $D_{[210]} \pi_{[100]}$ (1) & $D_{[110]} \rho_{[000]}$ (1) & ${D^*}_{[110]} \pi_{[100]}$ (2) & $D_{[211]} \pi_{[100]}$ (1) & $D_{[200]} {f_0}_{[000]}$ (1) \\
			$\bar{q} \mathbf{\Gamma} q$ (20) & ${D^*}_{[100]} \pi_{[100]}$ (2) & $D_{[100]} \eta_{[100]}$ (1) & $D_{[100]} {f_0}_{[100]}$ (1) & ${D^*}_{[111]} \pi_{[000]}$ (1) & $D_{[100]} \eta_{[110]}$ (1) & ${D^*}_{[100]} \pi_{[100]}$ (1) \\
			& ${D^*}_{[110]} \pi_{[000]}$ (1) & $D_{[110]} \rho_{[000]}$ (1) & ${D^*}_{[000]} \pi_{[110]}$ (1) & ${D^*}_{[111]} \eta_{[000]}$ (1) & $D_{[110]} \eta_{[100]}$ (1) & ${D^*}_{[200]} \pi_{[000]}$ (1) \\
			& ${D^*}_{[111]} \pi_{[100]}$ (2) & ${D_s}_{[100]} \bar{K}_{[100]}$ (1) & ${D^*}_{[100]} \pi_{[100]}$ (2) & ${D_s^*}_{[111]} \bar{K}_{[000]}$ (1) & $D_{[111]} \rho_{[000]}$ (1) & ${D^*}_{[210]} \pi_{[100]}$ (2) \\
			& ${D^*}_{[110]} \eta_{[000]}$ (1) & ${D^*}_{[000]} \pi_{[110]}$ (1) & ${D^*}_{[110]} \pi_{[000]}$ (1) & ${D_0}_{[111]} \pi_{[000]}$ (1) & $D_{[110]} {f_0}_{[100]}$ (1) & ${D^*}_{[100]} \eta_{[100]}$ (1) \\
			& ${D_s^*}_{[110]} \bar{K}_{[000]}$ (1) & ${D^*}_{[100]} \pi_{[100]}$ (1) & ${D^*}_{[111]} \pi_{[100]}$ (1) & $\bar{q} \mathbf{\Gamma} q$ (36) & ${D_s}_{[100]} \bar{K}_{[110]}$ (1) & ${D^*}_{[200]} \eta_{[000]}$ (1) \\
			& ${D_0}_{[110]} \pi_{[000]}$ (1) & ${D^*}_{[110]} \pi_{[000]}$ (1) & ${D^*}_{[110]} \eta_{[000]}$ (1) &  & ${D_s}_{[110]} \bar{K}_{[100]}$ (1) & ${D_s^*}_{[200]} \bar{K}_{[000]}$ (1) \\
			& $\bar{q} \mathbf{\Gamma} q$ (52) & ${D^*}_{[111]} \pi_{[100]}$ (2) & ${D_s^*}_{[110]} \bar{K}_{[000]}$ (1) &  & ${D^*}_{[100]} \pi_{[110]}$ (3) & $\bar{q} \mathbf{\Gamma} q$ (32) \\
			&  & ${D^*}_{[100]} \eta_{[100]}$ (1) & $\bar{q} \mathbf{\Gamma} q$ (52) &  & ${D^*}_{[110]} \pi_{[100]}$ (3) &  \\
			&  & ${D^*}_{[110]} \eta_{[000]}$ (1) &  &  & ${D^*}_{[111]} \pi_{[000]}$ (1) &  \\
			&  & ${D^*}_{[110]} {f_0}_{[000]}$ (1) &  &  & ${D^*}_{[111]} \eta_{[000]}$ (1) &  \\
			&  & ${D_s^*}_{[110]} \bar{K}_{[000]}$ (1) &  &  & ${D_s^*}_{[111]} \bar{K}_{[000]}$ (1) &  \\
			&  & $\bar{q} \mathbf{\Gamma} q$ (44) &  &  & $\bar{q} \mathbf{\Gamma} q$ (60) &  \\
			\bottomrule
		\end{tabular}
	}
	\caption{$I=1/2$ meson-meson-like operators used in the computation of the spectra presented in section~\ref{subsec:fv_spectra}. The subscripts in brackets indicate the lattice momentum of the single-meson operator. The number in round parentheses indicates how many distinct operators have been included for the given meson-meson combination and momenta.}
	\label{tab:app:dstarpi:ops2}
\end{table}

%% file: appendix_amp_variations.tex
\section{List of amplitude variations}
\label{sec:app:amps}

This appendix contains tables listing the amplitude-parameterisation variations based on the $K$-matrix formalism. The tables contain a label for the amplitude in the first column and the fit quality in terms of the reduced $\chi^2$ minimum in the last column. The other columns indicate which parameters in the general $K$-matrix formula given by equation~\ref{eq:kmat_gen} were allowed to float in the fit. Dashes mean that the parameter is fixed to zero whereas single letters indicate the polynomial order of the free parameter. 'c' means 'constant', 'l' means linear and 'q' quadratic. A lower-case letter means that only a single term of the given order has been included. A capital letter means that terms of all orders up to the indicated order are included.
 In every table, the first row is the reference parametrisation.

\FloatBarrier

\subsection{$J^P=1^+$ variations}

The amplitude variations are given in table~\ref{tab:amps_1p}. The first two groups of amplitudes correspond to variations of the $\tso$ and $\tdo$ partial waves that have no linear or higher polynomial dependence on $s$. The third group includes amplitudes with $\dstarpi\{\tdo\}$ contributions that have higher polynomial orders in $s$. The last group allows for $\dstarpi\{\tso\}$ coupling parameters that depend linearly on $s$. The channel labels are $1 \to \dstarpi\{\tso\}$, $2 \to \dstarpi\{\tdo\}$, $3 \to \dstareta\{\tso\}$ and $4 \to \dsstarkbar\{\tso\}$.

\newlength\lengtha \setlength\lengtha{4mm} 
\newlength\lengthb \setlength\lengthb{8mm}

\begin{table}[htb!]

	\resizebox{\textwidth}{!}{
	\begin{tabular}{c@{\hspace*{\lengtha}}cccc@{\hspace*{\lengthb}}cccc@{\hspace*{\lengthb}}cccccccccc@{\hspace*{\lengthb}}c}
		\toprule
		\multirow{2}{*}{Amplitude} & \multicolumn{4}{c}{$g_{0,i}$} &  \multicolumn{4}{c}{$g_{1,i}$} &  \multicolumn{10}{c}{$\gamma^{(n)}_{ij}$} & \multirow{2}{*}{ $\chi^2/N_{\text{DoF}}$} \\
		& 1 & 2 & 3 & 4 & 1 & 2 & 3 & 4 & 11 & 12 & 13 & 14 & 22 & 23 & 24 & 33 & 34 & 44 & \\
		\toprule
		\textbf{(1)} & c & - & c & c & c & c & - & - & c & - & - & - & - & - & - & - & - & c & $\tfrac{100.8}{107-17}$ \\
		(2) & c & c & c & c & c & c & - & - & c & - & - & - & - & - & - & - & - & - & $\tfrac{102.1}{107-17}$ \\
		(3) & c & c & - & c & c & c & - & - & c & - & - & - & - & - & - & c & - & c & $\tfrac{101.3}{107-18}$ \\
		(4) & c & c & c & c & - & c & - & - & c & - & - & - & - & - & - & - & - & c & $\tfrac{103.4}{107-17}$ \\
		(5) & c & c & c & c & c & c & - & c & c & - & - & - & - & - & - & - & - & c & $\tfrac{100.1}{107-19}$ \\
		(6) & c & c & c & c & c & c & - & - & c & - & - & - & - & - & c & - & - & c & $\tfrac{100.1}{107-19}$ \\
		(7) & c & c & c & c & c & c & - & - & c & c & - & - & - & - & - & - & - & c & $\tfrac{100.1}{107-19}$ \\
		(8) & c & c & c & c & c & c & c & - & c & - & - & - & - & - & - & - & - & c & $\tfrac{100.2}{107-19}$ \\
		(9) & c & c & c & c & c & c & - & - & c & - & - & - & - & c & - & - & - & c & $\tfrac{100.2}{107-19}$ \\
		(10) & c & c & c & c & c & c & - & - & c & - & - & c & - & - & - & - & - & c & $\tfrac{100.2}{107-19}$ \\
		(11) & c & c & c & c & c & c & - & - & c & - & - & - & - & - & - & c & - & c & $\tfrac{100.3}{107-19}$ \\
		(12) & c & c & c & c & c & c & - & - & c & - & - & - & - & - & - & - & c & c & $\tfrac{100.3}{107-19}$ \\
		(13) & c & c & c & c & c & c & - & - & c & - & c & - & - & - & - & - & - & c & $\tfrac{100.3}{107-19}$ \\
		(14) & c & c & c & - & c & c & - & - & c & - & - & c & - & - & - & - & c & c & $\tfrac{100.8}{107-19}$ \\
		\midrule
		(15) & c & c & c & c & c & - & - & - & c & - & - & - & - & - & - & - & - & c & $\tfrac{101.3}{107-17}$ \\
		(16) & c & c & c & c & c & c & - & - & c & - & - & - & - & - & c & - & - & c & $\tfrac{100.1}{107-19}$ \\
		(17) & c & c & c & c & c & c & - & - & c & c & - & - & - & - & - & - & - & c & $\tfrac{100.1}{107-19}$ \\
		(18) & c & c & c & c & c & c & - & - & c & - & - & - & - & c & - & - & - & c & $\tfrac{100.2}{107-19}$ \\
		(19) & c & c & c & c & c & c & - & - & c & - & - & - & c & - & - & - & - & c & $\tfrac{100.2}{107-19}$ \\
		\midrule
		\emph{(20)}* & c & - & c & c & c & c & - & - & c & - & - & - & l & - & - & - & - & c & $\tfrac{104.8}{107-18}$ \\
		\emph{(21)}* & c & - & c & c & c & c & - & - & c & - & - & - & L & - & - & - & - & c & $\tfrac{96.7}{107-19}$ \\
		\emph{(22)}* & c & - & c & c & c & c & - & - & c & - & - & - & L & - & - & - & - & c & $\tfrac{95.8}{107-20}$ \\
		\emph{(23)}* & c & - & c & c & c & c & - & - & c & - & - & - & Q & - & - & - & - & c & $\tfrac{95.3}{107-21}$ \\
		\midrule
		\emph{(24)}${}^\dagger$ & L & - & c & c & c & c & - & - & c & - & - & - & - & - & - & - & - & c & $\tfrac{100.3}{107-18}$ \\
		\emph{(25)}${}^\dagger$ & c & - & c & L & c & c & - & - & c & - & - & - & - & - & - & - & - & c & $\tfrac{105.5}{107-18}$ \\
		\bottomrule
	\end{tabular}
	}
	\\
	$*$ physical-sheet poles; only used in result for state ($\mathsf{II})$\\
	$\dagger$ not used in analysis

	\caption{Amplitude variations used in the $J^P = 1^+$ scattering analysis as described in section~\ref{subsec:variations}.}
	\label{tab:amps_1p}
\end{table}

\FloatBarrier
\subsection{$J^P=2^+$ variations}

The amplitude variations are given in table~\ref{tab:amps_2p}. The channel labels are $1 \rightarrow \dpi\{\odt\}$ and $2 \rightarrow \dstarpi\{\tdt\}$.

\begin{table}[htb!]
	\centering

	\begin{tabular}{c@{\hspace*{\lengtha}}cc@{\hspace*{\lengthb}}ccc@{\hspace*{\lengthb}}c}
			\toprule
			\multirow{2}{*}{Amplitude} & \multicolumn{2}{l}{$g_{0,i}$} &  \multicolumn{3}{c}{\hspace{-0.6cm}$\gamma_{ij}$} & \multirow{2}{*}{ $\chi^2/N_{\text{DoF}}$} \\
			 & 1 & 2  & 11 & 12 & 22 & \\
			\toprule
			\textbf{(1)} & c & c & c & - & - & $\tfrac{100.8}{107-17}$ \\
			\midrule
			(2) & c & - & c & - & c & $\tfrac{101.4}{107-17}$ \\
			(3) & c & c & - & - & - & $\tfrac{103.2}{107-17}$ \\
			(4) & c & c & c & c & - & $\tfrac{100.2}{107-19}$ \\
			\bottomrule
	\end{tabular}

\caption{Amplitude variations in $J^P = 2^+$ scattering analysis}
\label{tab:amps_2p}
\end{table}

\FloatBarrier

\subsection{Negative-parity variations}

In the reference parametrisation, $J^P = 0^-$ is fixed to zero. As a variation, we include a single parametrisation that allows for a constant $\gamma_{\tpz}$ while all other parameters correspond to the baseline parametrisation. The fit yields
\begin{equation}
	\gamma_{\tpz} = 20 \pm 53
\end{equation}
with
\begin{align}
	\chi^2/ N_\mathrm{dof} = \tfrac{100.14}{107-19} = 1.14\,.
\end{align}
The $J^P = 1^-$ variations are given in table~\ref{tab:amps_1m}. Channel labels are $1 \rightarrow \dpi\{\opo\}$ and $2 \rightarrow \dstarpi\{\tpo\}$.

\begin{table}[htb!]
	\centering
	
		\begin{tabular}{c@{\hspace*{\lengthb}}ccc@{\hspace*{\lengthb}}c}
			\toprule
			\multirow{2}{*}{Amplitude} &  \multicolumn{3}{c}{\hspace{-0.3cm}$\gamma^{(n)}_{ij}$} & \multirow{2}{*}{ $\chi^2/N_{\text{DoF}}$} \\
			  & 11 & 12 & 22 &  \\
			\toprule
			\textbf{(1)} & c & c & c & $\tfrac{100.8}{107-17}$ \\
			\midrule
			(2) & c & - & c & $\tfrac{104.2}{107-17}$ \\
			(3) & L & c & c & $\tfrac{104.6}{107-19}$ \\
			(4) & c & L & c & $\tfrac{105.4}{107-19}$ \\
			\bottomrule
		\end{tabular}
	
	\caption{Amplitude variations in $J^P = 1^-$ scattering analysis}
	\label{tab:amps_1m}
\end{table}
For $J^P = 2^-$, the reference parametrisation includes a constant K matrix (c.f. equation~\ref{eq:kmat_fit_res}). As a variation, we include a single parametrisation where $\gamma_{\tpt}$ is fixed to zero. This fit yields
\begin{align}
	\chi^2/ N_\mathrm{dof} = \tfrac{101.55 }{107-17} = 1.13\,.
\end{align}

\subsection{Fits with a reduced dataset}

In sections~\ref{subsec:variations} and \ref{subsec:poles_1p_narrow} we compare to a set of fits based on a reduced dataset, which excludes levels beyond $E_\dstareta |_\thr$ in irreps with a $J^P=1^+$ subduction. These amplitudes and the corresponding fit results are given in table~\ref{tab:elastic_amps}. The channel labels are $1 \rightarrow \dstarpi\{\tso\}$, $2 \rightarrow \dstarpi\{\tdo\}$, $3 \rightarrow \dstarpi\{\tpz\}$ and $4 \rightarrow \dstarpi\{\tpt\}$. The fits do not enter in any of the final results.

\begin{table}[htb!]
	\centering

\begin{tabular}{c@{\hspace*{\lengtha}}cc@{\hspace*{\lengthb}}cc@{\hspace*{\lengthb}}ccc@{\hspace*{\lengthb}}c@{\hspace*{\lengthb}}c@{\hspace*{\lengthb}}c}
	\toprule
	\multirow{2}{*}{Amplitude} & \multicolumn{2}{c}{$g_{0,i}$} &  \multicolumn{2}{c}{$g_{1,i}$} &  \multicolumn{5}{c}{$\gamma^{(n)}_{ij}$} & \multirow{2}{*}{ $\chi^2/N_{\text{DoF}}$} \\
	& 1 & 2 & 1 & 2 & 11 & 12 & 22 & 33 & 44 & \\
	\toprule
	
	(1) & c & - & c & c & c & - & c & - & - & $\tfrac{100.0}{99-14}$ \\
	(2) & c & c & c & - & c & - & c & - & - & $\tfrac{100.0}{99-14}$ \\
	(3) & c & c & c & c & c & - & c & - & - & $\tfrac{98.5}{99-15}$ \\
	(4) & c & - & c & c & c & - & c & - & - & $\tfrac{100.6}{99-14}$ \\
	(5) & c & - & c & c & c & c & c & - & - & $\tfrac{98.6}{99-15}$ \\
	(6) & c & - & c & c & c & - & c & c & - & $\tfrac{99.2}{99-15}$ \\
	(7) & c & - & c & c & c & - & c & - & c & $\tfrac{99.6}{99-15}$ \\
	
	\bottomrule
\end{tabular}

\caption{Amplitude variations excluding the $\dstareta\{\tso\}$ and $\dsstarkbar\{\tso\}$ channels, fitted to a reduced dataset of 99 energy levels, excluding the highest levels in $[000] T_1^+$. This is described in sections~\ref{subsec:variations} and \ref{subsec:poles_1p_narrow}.}
\label{tab:elastic_amps}
\end{table}

\newpage
\subsection{Hadron-mass and anisotropy variations}

The anisotropy and stable hadron masses (table~\ref{tab:dstarpi:had_masses}) are varied one by one within their respective uncertainties and in each case a fit is performed using the reference parametrisation (equation~\ref{eq:kmat_coupled}). The corresponding fit results are shown in table~\ref{tab:syst_fits}. The value and error of the anisotropy is $\xi = 3.444 \pm 0.053$.

\begin{table}[htb!]
	\centering
	
		\begin{tabular}{cc}
			\toprule
			Amplitude & $\chi^2/N_{\text{DoF}}$ \\
			\toprule
			 $\xi \uparrow$ & $\tfrac{101.2}{107-17}$ \\
			$\xi \downarrow$ & $\tfrac{100.6}{107-17}$ \\
			$m_{\pi} \uparrow$ & $\tfrac{101.8}{107-17}$ \\
			$m_{\pi} \downarrow$ & $\tfrac{100.0}{107-17}$ \\
			$m_{\eta} \uparrow$ & $\tfrac{100.8}{107-17}$ \\
			$m_{\eta} \downarrow$ & $\tfrac{100.9}{107-17}$ \\
			$m_{\bar{K}} \uparrow$ & $\tfrac{100.8}{107-17}$ \\
			$m_{\bar{K}} \downarrow$ & $\tfrac{100.9}{107-17}$ \\
			$m_{D} \uparrow$ & $\tfrac{103.4}{107-17}$ \\
			$m_{D} \downarrow$ & $\tfrac{99.5}{107-17}$ \\
			$m_{D^*} \uparrow$ & $\tfrac{104.0}{107-15}$ \\
			$m_{D^*} \downarrow$ & $\tfrac{101.5}{107-17}$ \\
			\bottomrule
		\end{tabular}
	
	\caption{Fit results from hadron-mass and anisotropy variations}
	\label{tab:syst_fits}
\end{table}